\documentclass[%
  reprint,
  showpacs,preprintnumbers,
  amsmath,amssymb,
  aps,
  prb,
]{revtex4-2}

\usepackage{graphicx}[]
\usepackage{dcolumn}
\usepackage{bm}
\usepackage{multirow}
\usepackage{braket}
\usepackage{color}
\usepackage{ulem}

\usepackage{url}

\begin{document}

\preprint{APS/123-QED}

\title{
Multifarious skyrmion phases on a trilayer triangular lattice
}

\author{Satoru Hayami}
\affiliation{
Department of Applied Physics, The University of Tokyo, Tokyo 113-8656, Japan
}
 
\begin{abstract}
The instability toward a magnetic skyrmion crystal in centrosymmetric trilayer magnets is investigated based on a spin model with layer-dependent Dzayloshinskii-Moriya interaction. 
We find various types of skyrmion crystal phases with different skyrmion numbers in a low-temperature phase diagram by performing the simulated annealing. 
In addition to the N\'eel skyrmion crystal phase that is expected to emerge in the presence of the polar-type Dzayloshinskii-Moriya interaction, we obtain the skyrmion crystal phases characteristics of the layered system: the twisted surface skyrmion crystal, anti-skyrmion crystal, and high-topological-number skyrmion crystal phases. 
The rich magnetic phases are brought about by the synergy among the layer-dependent Dzayloshinskii-Moriya interaction, interlayer exchange interaction, and an external magnetic field. 
Our results indicate that the layer degree of freedom at the surface and heterostructures provides a good platform to engineer and design the topological spin textures. 
\end{abstract}
\maketitle

\section{Introduction}

Spiral magnetism has long been studied in condensed matter physics, as it manifests itself not only in unusual magnetism but also in peculiar transport and multiferroics  phenomena~\cite{Kaplan_PhysRev.116.888, yoshimori1959new, VILLAIN1959303, Kaplan_PhysRev.124.329, Elliott_PhysRev.124.346}.
The concept of spiral magnetism brings about novel phases of matter, such as spiral spin liquids~\cite{bergman2007order,Zaharko_PhysRevB.84.094403,Gao2016Spiral,Iqbal_PhysRevB.98.064427,Buessen_PhysRevLett.120.057201,Pohle_PhysRevB.104.024426}, skyrmion crystals (SkXs)~\cite{Muhlbauer_2009skyrmion,yu2010real,nagaosa2013topological, fert2017magnetic, kurumaji2019skyrmion, Tokura_doi:10.1021/acs.chemrev.0c00297,hayami2021topological}, hedgehog lattices~\cite{tanigaki2015real,kanazawa2017noncentrosymmetric,fujishiro2019topological,Kanazawa_PhysRevLett.125.137202,grytsiuk2020topological,Ishiwata_PhysRevB.101.134406,Okumura_PhysRevB.101.144416,Mendive-Tapia_PhysRevB.103.024410,Shimizu_PhysRevB.103.054427,Aoyama_PhysRevB.103.014406}, meron-antimeron crystals~\cite{Lin_PhysRevB.91.224407,yu2018transformation,hayami2018multiple,Hayami_PhysRevB.104.094425,chen2022triple}, and chiral stripes~\cite{Solenov_PhysRevLett.108.096403,Ozawa_doi:10.7566/JPSJ.85.103703,Shimokawa_PhysRevB.100.224404,yambe2020double}. 
Moreover, as the formation of the spiral spin texture often leads to the breaking of spatial inversion symmetry, it becomes a source of parity-violating electronic states and physical phenomena, such as an antisymmetric spin-split band structure~\cite{Hayami_PhysRevB.101.220403,Hayami_PhysRevB.102.144441,Hayami_PhysRevB.105.024413}, magnetoelectric effect~\cite{Katsura_PhysRevLett.95.057205,Mostovoy_PhysRevLett.96.067601,cheong2007multiferroics,Bulaevskii_PhysRevB.78.024402,seki2012observation,white2012electric,tokura2014multiferroics,mochizuki2015dynamical,Gobel_PhysRevB.99.060406,kurumaji2020spiral}, and nonreciprocal transport~\cite{Miyahara_JPSJ.81.023712,Miyahara_PhysRevB.89.195145,Seki_PhysRevB.93.235131,giordano2016spin,tokura2018nonreciprocal,yokouchi2018current,Hoshino_PhysRevB.97.024413,Baryakhtar_PhysRevB.99.104407,okumura2019spin,Santos_PhysRevB.102.104401,seki2020propagation,hayami2021phase,Hayami2022_2}. 
In this way, spiral spin ordering provides a rich playground for exploring intriguing functional materials that might be utilized for future electronic and spintronic device applications. 

The realization of the spiral ordering has been achieved under several different spin interactions: the Dzyaloshinskii-Moriya (DM) interaction in noncentrosymmetric magnets~\cite{dzyaloshinsky1958thermodynamic,moriya1960anisotropic}, the frustrated exchange interaction in insulating magnets, and the Ruderman-Kittel-Kasuya-Yosida (RKKY) interaction in itinerant magnets~\cite{Ruderman, Kasuya, Yosida1957}.  
Remarkably, these interactions also induce instabilities toward different types of multiple-$Q$ states, which are represented by a superposition of spiral waves. 
We here introduce instabilities toward the SkXs and their different aspects under each interaction. 
In the case of the DM interaction, the SkX is stabilized when applying an external magnetic field~\cite{rossler2006spontaneous, Yi_PhysRevB.80.054416, Binz_PhysRevLett.96.207202, Binz_PhysRevB.74.214408}. 
In this case, the helicity and vorticity of the SkX are determined by the DM vector built in the lattice structure; the Bloch SkX, N\'eel SkX, and anti-SkX emerge in the polar, chiral, and $D_{2d}$ systems~\cite{nagaosa2013topological, Tokura_doi:10.1021/acs.chemrev.0c00297}. 
As the spiral modulation period is determined by the ratio of the DM interaction and the ferromagnetic exchange interaction, the SkX usually exhibits the long-period structure, although the recent studies have shown that the short-period SkX can be engineered by taking into account the multi-spin interaction~\cite{heinze2011spontaneous,brinker2019chiral,paul2020role, Brinker_PhysRevResearch.2.033240,hayami2021field}, anisotropic exchange interaction~\cite{Hayami_PhysRevLett.121.137202, Kato_PhysRevB.104.224405}, and spin-charge-coupled intearction~\cite{Mohanta_PhysRevB.100.064429,Kathyat_PhysRevB.103.035111}. 
More recently, it was shown that the interlayer exchange interaction in the nonsymmorphic lattice system with a screw axis also leads to the SkX~\cite{Hayami2022_4}. 

In contrast to the DM-interaction mechanism, the latter two mechanisms based on the frustrated exchange interaction and the RKKY interaction do not require the noncentrosymmetric lattice structures; they give rise to the SkXs in centrosymmetric lattice systems~\cite{batista2016frustration,hayami2021topological}. 
For the mechanism by the frustrated exchange interaction, the SkX appears by incorporating the effect of thermal fluctuations~\cite{Okubo_PhysRevLett.108.017206}, uniaxial spin anisotropy~\cite{leonov2015multiply,Lin_PhysRevB.93.064430,Hayami_PhysRevB.93.184413,Lin_PhysRevLett.120.077202,Lohani_PhysRevX.9.041063,Hayami_PhysRevB.103.224418, Hayami2022_3}, and nonmangetic impurities~\cite{Hayami_PhysRevB.94.174420}. 
Similarly, the studies have revealed that the SkX in itienrant magnets is stabilized by considering thermal fluctuations~\cite{Mitsumoto_PhysRevB.104.184432,mitsumoto2021skyrmion,kato2022magnetic}, single-ion anisotropy~\cite{Wang_PhysRevLett.124.207201}, spin-charge coupling including the biquadratic interaction~\cite{Ozawa_PhysRevLett.118.147205,Hayami_PhysRevB.95.224424,Hayami_PhysRevB.99.094420,hayami2020multiple,Hayami_PhysRevResearch.3.043158,Hayami_10.1088/1367-2630/ac3683,wang2021skyrmion}, and circularly polarized microwave field~\cite{Eto_PhysRevB.104.104425} in addition to the RKKY interaction. 
In these cases, the helicity and vorticity of the SkX are arbitrary, which results in peculiar symmetry breaking states~\cite{Okubo_PhysRevLett.108.017206,Capic_PhysRevResearch.1.033011,mitsumoto2021skyrmion} and transport properties~\cite{Lin_PhysRevB.93.064430,zhang2017skyrmion,Xia_PhysRevApplied.11.044046,zhang2020skyrmion,Zhang_PhysRevB.101.144435,wang2021stimulated,Yao_PhysRevB.105.014444}. 
In centrosymmetric magnets, the degeneracy in the SkXs is (partly) lifted in the presence of the symmetric anisotropic exchange interaction and dipolar interaction, the former of which arises from the discrete lattice symmetry~\cite{amoroso2020spontaneous,Hayami_doi:10.7566/JPSJ.89.103702, Hayami_PhysRevB.103.054422, Wang_PhysRevB.103.104408,Hayami2022_1,yambe2022effective}. 
These anisotropic interactions manifest themselves in the stabilization of the SkXs~\cite{Hayami_PhysRevB.103.024439, Wang_PhysRevB.103.104408, Utesov_PhysRevB.103.064414, amoroso2020spontaneous,gao2020fractional, yambe2021skyrmion, amoroso2021tuning,hayami2022multiple,utesov2021mean,yambe2022effective}, which might be important to reproduce the experimental phase diagrams in SkX-hosting materials, Gd$_3$Ru$_4$Al$_{12}$~\cite{hirschberger2019skyrmion,Hirschberger_10.1088/1367-2630/abdef9}, GdRu$_2$Si$_2$~\cite{khanh2020nanometric,Yasui2020imaging, khanh2022zoology}, and EuAl$_4$~\cite{Shang_PhysRevB.103.L020405, kaneko2021charge,Zhu2022, takagi2022}. 

In the present study, we investigate another stabilization mechanism of the SkX in centrosymmetric systems by focusing on the layer degree of freedom. 
Especially, we examine the effect of the layer-dependent polar DM interaction that originates from the local inversion symmetry breaking~\cite{zhang2014hidden,Hayami_PhysRevB.90.024432,Fu_PhysRevLett.115.026401,Razzoli_PhysRevLett.118.086402,hayami2016emergent,gotlieb2018revealing,Huang_PhysRevB.102.085205,Ishizuka_PhysRevB.98.224510}. 
It has been recently clarified that a staggered DM interaction yields the SkX in centrosymmetric bilayer magnets~\cite{Hayami_PhysRevB.105.014408,lin2021skyrmion}. 
We here extend such studies to a trilayer system, where a middle layer without the DM interaction is sandwiched by the upper and lower layers with the opposite sign of the DM vectors. 
By performing the simulated annealing for the spin model with the layer-dependent DM interaction, we show that multifarious SkXs with different layer-dependent skyrmion numbers are stabilized at low temperatures as a consequence of the interplay between an interlayer exchange interaction and an external magnetic field. 
We find that the skyrmion number in the middle layer is sensitive to the change in the model parameters. 
Notably, we discover the anti-SkX for the small interlayer exchange interaction, which has not been usually stabilized in the presence of the polar/chiral DM vector in the threefold-symmetric hexagonal and trigonal systems. 
We also find a twisted surface SkX and high-topological-number SkX in a wide range of model parameters. 
Our result provides a possibility of engineering the SkXs that is difficult to realize for bulk by using the layer, surface, and domain structures. 

The organization of this paper is as follows. 
In Sec.~\ref{sec: Model and method}, we introduce a spin model in the trilayer system with the layer-dependent DM interaction. 
We also present numerical methods based on the simulated annealing. 
In Sec.~\ref{sec: Result}, we examine the instability toward the SkX by constructing the low-temperature phase diagram. 
We discuss the nature of each phase obtained by the simulated annealing one by one. 
Section~\ref{sec: Summary} is devoted to a summary of the present paper. 
In Appendix~\ref{sec: app}, we present the result for different values of the DM interaction.

\section{Model and method}
\label{sec: Model and method}

\begin{figure}[t!]
\begin{center}
\includegraphics[width=1.0 \hsize ]{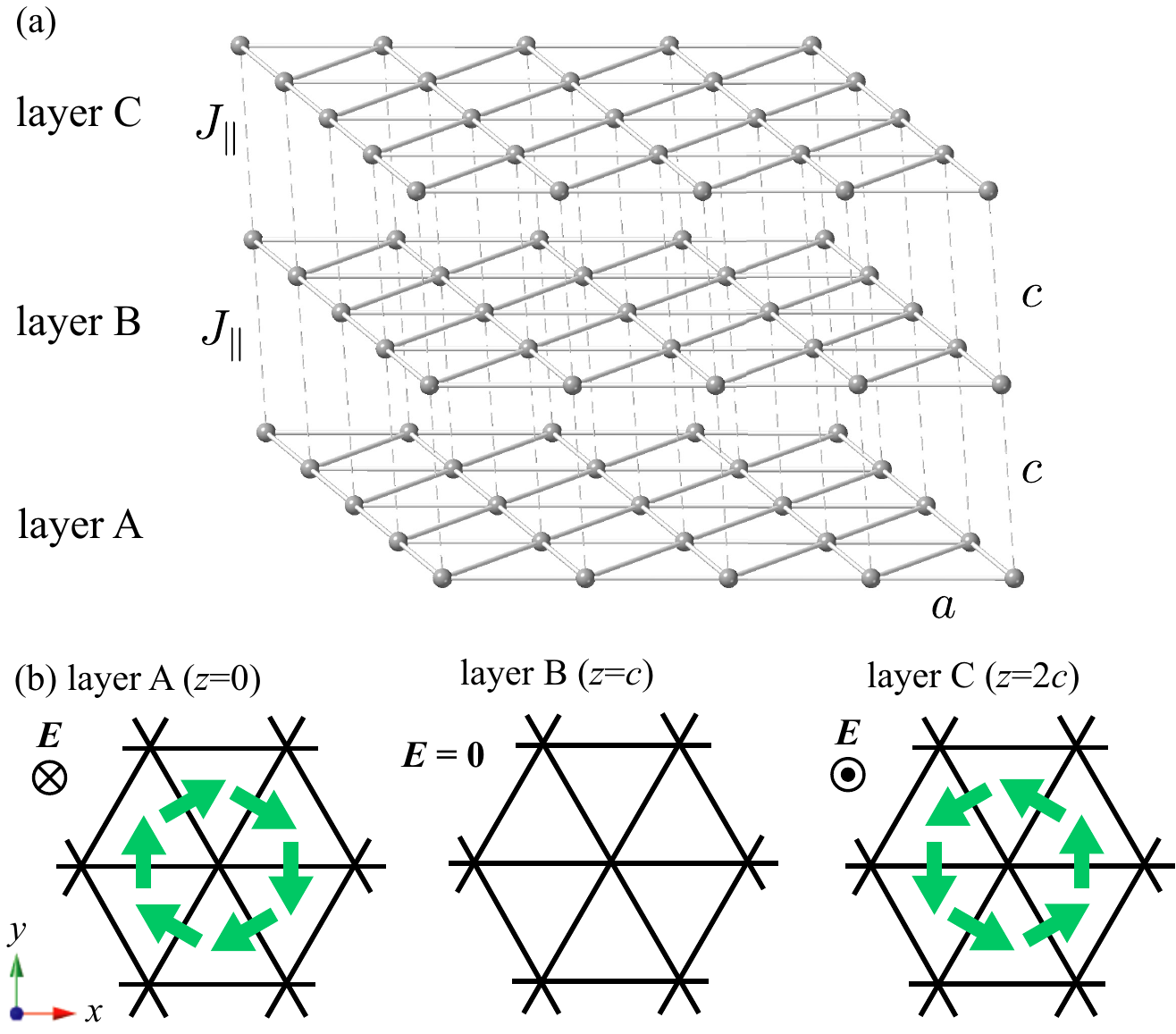} 
\caption{
\label{fig: lattice}
(a) The trilayer triangular-lattice system consisting of layers A, B, and C. 
The lattice constants, $a$ and $c$, and the interlayer exchange coupling $J_{\parallel}$ are also shown. 
(b) Three layers viewed from the $z$ axis. 
The green arrows denote the DM vectors in each layer, where $\bm{E}$ represents the local crystalline electric field. 
}
\end{center}
\end{figure}

We consider a trilayer triangular-lattice system as shown in Fig.~\ref{fig: lattice}(a); three triangular planes with the lattice constant $a$ lie in the $xy$ plane and they are coupled along the $z$ direction separated by a distance $c$.  
We label the lower, middle, and upper layers as layer A, layer B, and layer C, respectively. 
Three layers are not equivalent from the symmetry viewpoint; there is no inversion center for layers A and C, while there is an inversion center for layer B. 
Specifically, the point-group symmetry for layers A and C corresponds to $C_{6v}$, while that for layer B is $D_{6h}$. 
Reflecting such a difference, a local polar-type crystalline electric field is present only for layers A and C, which results in the layer-dependent DM interaction. 
We show the DM vectors in each layer in Fig.~\ref{fig: lattice}(b), whose directions are perpendicular to the intralayer bond direction and the $z$ direction. 
Owing to the inversion center at layer B, the direction of the DM vector for layers A and C are opposite and their magnitude is equivalent. 
In the following, we set $a=c=1$. 

The classical spin model in the trilayer system to incorporate the effect of the layer-dependent DM interaction is given by 
\begin{align}
\label{eq: Ham}
\mathcal{H}=\sum_{\eta}\mathcal{H}^{\perp}_{\eta}+\mathcal{H}^{\parallel}+\mathcal{H}^{{\rm Z}}, 
\end{align}
where the Hamiltonian is devided into three parts: the intralayer contribution $\mathcal{H}^{\perp}_{\eta}$ for layer $\eta=$ A, B, and C, the interlayer Hamiltonian $\mathcal{H}^{\parallel}$, and the Zeeman Hamiltonian $\mathcal{H}^{{\rm Z}}$, which are explictily shown as 
\begin{align}
\label{eq: Ham_perp}
\mathcal{H}^{\perp}_{\eta}=&  -\sum_{i,j} \left[ J_{ij} \bm{S}_{i} \cdot \bm{S}_{j} +  \bm{D}_{ij}^{(\eta)} \cdot  (\bm{S}_{i} \times \bm{S}_{j}) \right], \\
\label{eq: Ham_parallel}
\mathcal{H}^{\parallel}=& J_{\parallel} \sum_{i, \delta=\pm 1} \bm{S}_i \cdot \bm{S}_{i+\delta\hat{z}},\\
\label{eq: Ham_Zeeman}
\mathcal{H}^{{\rm Z}}=&-H \sum_i S_i^z. 
\end{align}
The intralayer Hamiltonian $\mathcal{H}^{\perp}_{\eta}$ in Eq.~(\ref{eq: Ham_perp}) consists of the layer-independent exchange interaction $J_{ij}$ and the layer-dependent DM interaction to satisfy $\bm{D}_{ij}^{({\rm A})}=-\bm{D}_{ij}^{({\rm C})}$ and $\bm{D}_{ij}^{({\rm B})}=0$ from the symmetry consideration; we set $|\bm{D}_{ij}^{({\rm A})}|=|\bm{D}_{ij}^{({\rm C})}|=D_{ij}$. 
As discussed above, the DM vectors are set by the green arrows in Fig.~\ref{fig: lattice}(b).
The interlayer Hamiltonian $\mathcal{H}^{\parallel}$ in Eq.~(\ref{eq: Ham_parallel}) represents the exchange coupling between the nearest-neighbor spins along the $z$ direction with the coupling constant $J_{\parallel}$; $J_{\parallel}>0$ ($J_{\parallel}<0$) stands for the antiferromagnetic (ferromagnetic) exchange interaction. 
The Zeeman Hamiltonian $\mathcal{H}^{{\rm Z}}$ in Eq.~(\ref{eq: Ham_Zeeman}) describes the Zeeman coupling to an external magnetic field along the $z$ direction. 
In the model in Eq.~(\ref{eq: Ham}), we neglect a long-range dipole-dipole interaction, which can affect the SkX instability, for simplicity~\cite{Utesov_PhysRevB.103.064414, utesov2021mean}. 

When $J_{\parallel}=0$, the system is decoupled into three independent layers. 
For layers A and C, the magnetic phases while changing $H$ are similar, which are determined by the interplay between the exchange interaction $J_{ij}$ and the DM interaction $D_{ij}$. 
When we consider the ferromagnetic exchange interaction $J_{ij}>0$ and the interactions are limited to the nearest-neighbor ones, the single-$Q$ spiral state, the N\'eel SkX, and the fully-polarized state are stabilized against $H$, where the N\'eel SkX is described by a superposition of three spiral waves connected by threefold rotation~\cite{Yi_PhysRevB.80.054416, Mochizuki_PhysRevLett.108.017601, Rowland_PhysRevB.93.020404}. 
The difference between layers A and C appears in the helicity of the spiral and SkX phases due to the opposite sign of the DM interaction. 
On the other hand, for layer B, there is no instability toward the SkX against $H$ at low temperatures; the ferromagnetic state is stabilized for the ferromagnetic exchange interaction, or the conical spiral state, whose spiral plane lies on the $xy$ plane, is stabilized by considering the effect of further-neighbor antiferromagnetic exchange interactions in addition to the nearest-neighbor ferromagnetic exchange interaction. 

Starting from the above situation at $J_{\parallel}=0$, we investigate the effect of the interlayer exchange coupling $J_{\parallel}$ on magnetic phases. 
Especially, we focus on the possibility of the SkXs in the centrosymmetric system consisting of the different layers with the layer-dependent DM interaction in Figs.~\ref{fig: lattice}(a) and \ref{fig: lattice}(b). 
Owing to the opposite sign of the DM interaction for the upper and lower layers, there is a magnetic frustration for the middle layer, which might be a source of inducing nontrivial topological spin textures that is difficult to realize in the single-layer system. 
Similar analyses have been recently performed for the bilayer system where the DM interaction is present in both layers with the opposite sign~\cite{hitomi2014electric,hitomi2016electric,yatsushiro2020odd,Yatsushiro_PhysRevB.102.195147}; the SkX is robustly stabilized by the interplay between the staggered DM interaction and interlayer exchange interaction even in the centrosymmetric lattice structure~\cite{Hayami_PhysRevB.105.014408,lin2021skyrmion}. 
The present trilayer model is regarded as an extension of the bilinear model. 
Meanwhile, the present trilayer system is qualitatively different from the bilayer system since there is no DM interaction in the middle layer (layer B) owing to the presence of the inversion center in the middle layer rather than the center of the bond between the adjacent layers~\cite{Maruyama_doi:10.1143/JPSJ.81.034702}. 
Reflecting such a difference, the trilayer model exhibits a rich phase diagram regarding the SkXs compared to the bilayer model, as shown in Sec.~\ref{sec: Result}. 

To investigate the SkX instability in such a layered system, we examine the competition between the interlayer exchange interaction and the layer-dependent DM interaction. 
For that purpose, we ignore spatial fluctuations of spins in the $xy$ plane by simplifying the intralayer Hamiltonian in Eq.~(\ref{eq: Ham_perp}) as 
\begin{align}
\label{eq:Ham_perp2}
\tilde{\mathcal{H}}^{\perp}_\eta=&  -\sum_{\nu}  \Big[ J \bm{S}^{(\eta)}_{\bm{Q}_{\nu}} \cdot \bm{S}^{(\eta)}_{-\bm{Q}_{\nu}}+ i   \bm{D}^{(\eta)}_\nu \cdot ( \bm{S}^{(\eta)}_{\bm{Q}_{\nu}} \times \bm{S}^{(\eta)}_{-\bm{Q}_{\nu}}) \Big],  
\end{align}
where $\bm{S}^{(\eta)}_{\bm{Q}_{\nu}}$ is the Fourier transform of $\bm{S}_i$ with wave vector $\bm{Q}_\nu$ for layer $\eta=$ A, B, and C; the subscript $\nu$ represents the index of the wave vectors. 
In Eq.~(\ref{eq:Ham_perp2}), we only consider the dominant $\bm{q}$ contributions that give global energy minima in momentum space, which is obtained by evaluating the Fourier transform of $\mathcal{H}^{\perp}_{\eta}$ in Eq.~(\ref{eq: Ham_perp}). 
Owing to the sixfold rotational symmetry of the triangular lattice, global minima appear, at least, at six wave vectors except for high-symmetric wave vectors, such as $\bm{q}=\bm{0}$ and the Brillouinze zone boundary~\cite{leonov2015multiply, Hayami_PhysRevB.103.224418, Hayami_PhysRevB.105.014408}. 
We suppose global minima at $\bm{Q}_1=Q(1,0)$, $\bm{Q}_2=Q(-1/2,\sqrt{3}/2)$, $\bm{Q}_3=Q(-1/2,-\sqrt{3}/2)$, $\bm{Q}_4=-\bm{Q}_1$, $\bm{Q}_5=-\bm{Q}_2$, and $\bm{Q}_6=-\bm{Q}_3$ with $Q=\pi/3$; $J \equiv J^{(\eta)}_{\bm{Q}_\nu}$ and $\bm{D}^{(\eta)}_\nu \equiv \bm{D}^{(\eta)}_{\bm{Q}_\nu}$ similar to Ref.~\cite{Hayami_PhysRevB.105.014408}. 
Although the interactions at the higher-harmonic wave vectors like $\bm{Q}_1+\bm{Q}_2$ also contribute to the energy in the multiple-$Q$ states including the SkX~\cite{Hayami_doi:10.7566/JPSJ.89.103702, hayami2022multiple}, we neglect them by assuming their contribution is much smaller than that at $\bm{Q}_\nu$. 
Furthermore, we drop off the contributions from the other $\bm{q}$ components in the interactions. 
Such a simplification is justified when considering the low-temperature phase diagram, where the $\bm{q}$-space dispersion is not important~\cite{leonov2015multiply}. 
In the end, we analyze the total Hamiltonian as follows: 
\begin{align}
\label{eq: Ham2}
\mathcal{H}=\sum_{\eta}\tilde{\mathcal{H}}^{\perp}_{\eta}+\mathcal{H}^{\parallel}+\mathcal{H}^{{\rm Z}}. 
\end{align}
Hereafter, we set $J=1$ as the energy unit of the model in Eq.~(\ref{eq: Ham2}). 
We choose the magnitude of the DM interaction as $|\bm{D}^{(\gamma)}_{\bm{Q}_\nu}|=D=0.2$ so that the SkX is stabilized under the magnetic field $H$ for layers A and C in the case of $J_{\parallel}=0$. 
In this situation, we investigate the instability toward the SkX while changing $J_{\parallel}$ and $H$. 
We also present the results for different $D$ in Appendix~\ref{sec: app}.

We construct the magnetic phase diagram by performing the simulated annealing for the model in Eq.~(\ref{eq: Ham2}) on the trilayer triangular lattice. 
The simulations are carried out with standard Metropolis local updates in real space following the manner in Ref.~\cite{Hayami_PhysRevB.105.014408}. 
Starting from a random spin configuration at high temperatures, we gradually reduce the temperature with a rate $T_{n+1}=\alpha T_n$, where $T_n$ is the temperature in the $n$th step. 
We set the initial temperature $T_0=0.1$-$1.0$ and the coefficient of geometrical cooling $\alpha=0.99999-0.999995$. 
The final temperature is set at $T=0.001$. 
We perform $10^5$-$10^6$ Monte Carlo sweeps for measurements at the final temperature. 
In addition to a random spin configuration, the simulations are performed from the spin configurations obtained at low temperatures when determining the phase boundaries in the phase diagram.
We adopt the periodic (open) boundary condition for the inplane ($z$) directions. 
The total number of spins is taken at $N=3\times L^2$ with $L= 48$. 
We confirmed that qualitative features are unchanged for different system sizes, such as $L=72$ and $96$.

We calculate the following quantities to identify the magnetic phases. 
The spin structure factor for layer $\eta$ is computed by 
\begin{align}
\label{eq:Sq}
S_{\eta}^\alpha(\bm{q})= \frac{1}{L^2} \sum_{i,j \in \eta} S^{\alpha}_i S^{\alpha}_j e^{i\bm{q}\cdot (\bm{r}_i-\bm{r}_j)}, 
\end{align}
for $\alpha=x,y,z$. 
We also compute $S_{\eta}^{xy}(\bm{q})=S_{\eta}^x(\bm{q})+S_{\eta}^y(\bm{q})$.
The net magnetization for each layer is given by
\begin{align}
M^\alpha_{\eta}=\frac{1}{L^2}\sum_{i \in \eta}S^{\alpha}_{i}.
\end{align}

The spin scalar chirality is represented by 
\begin{align}
\label{eq: chirality}
\chi^{\rm sc}_{\eta} &= \frac{1}{L^2} \sum_{\bm{R}\in \eta}\chi_{\bm{R}}\\
\chi_{\bm{R}}&= \bm{S}_{i} \cdot (\bm{S}_j \times \bm{S}_k),
\end{align}
where $\chi_{\bm{R}}$ represents the local scalar chirality at the position vector $\bm{R}$, which lies at the centers of upward and downward triangles with the vertices $i$, $j$, and $k$ in the counterclockwise order; the upward and downward triangles form the honeycomb network. 
Nonzero total scalar chirality $\chi^{\rm sc}=\chi^{\rm sc}_{\rm A}+ \chi^{\rm sc}_{\rm B}+ \chi^{\rm sc}_{\rm C}$ in the system is the origin of the topological Hall effect. 
We also compute the skyrmion number in each layer, which is given by 
\begin{align}
\label{eq:nsk}
n^{(\eta)}_\mathrm{sk} = \frac{1}{4\pi N_\mathrm{m}}\left\langle \sum_{\bm{R}\in \eta} \Omega^{\eta}_{\bm{R}} \right\rangle,
\end{align}
where $N_{\rm m}$ is number of the magnetic unit cell and $\Omega^{\eta}_{\bm{R}}$ is the skyrmion density for layer $\eta$~\cite{BERG1981412}: 
\begin{align}
\tan\left(\frac{\Omega^{\eta}_{\bm{R}}}{2}\right) = \frac{\bm{S}_i\cdot( \bm{S}_j\times \bm{S}_k)}{1+\bm{S}_i\cdot\bm{S}_j+\bm{S}_j\cdot\bm{S}_k+\bm{S}_k\cdot\bm{S}_i} .
\end{align}
For example, $n^{(\eta)}_\mathrm{sk}=-1$ when the N\'eel SkX appears for layer $\eta$, while $n^{(\eta)}_\mathrm{sk}=+1$ when the anti-SkX is realized. 
In the real-space picture, the SkX with $n^{(\eta)}_\mathrm{sk}=-1$ shows vortex-like winding of spins around the skyrmion core, while anti-SkX with $n^{(\eta)}_\mathrm{sk}=+1$ shows anti-vortex-like winding of spins. 
The averaged skyrmion number in the system is $n^{\rm ave}_{\rm sk}=(n^{\rm (A)}_\mathrm{sk}+n^{\rm (B)}_\mathrm{sk}+n^{\rm (C)}_\mathrm{sk})/3$.

\section{Results}
\label{sec: Result}

\begin{figure}[ht!]
\begin{center}
\includegraphics[width=0.8 \hsize ]{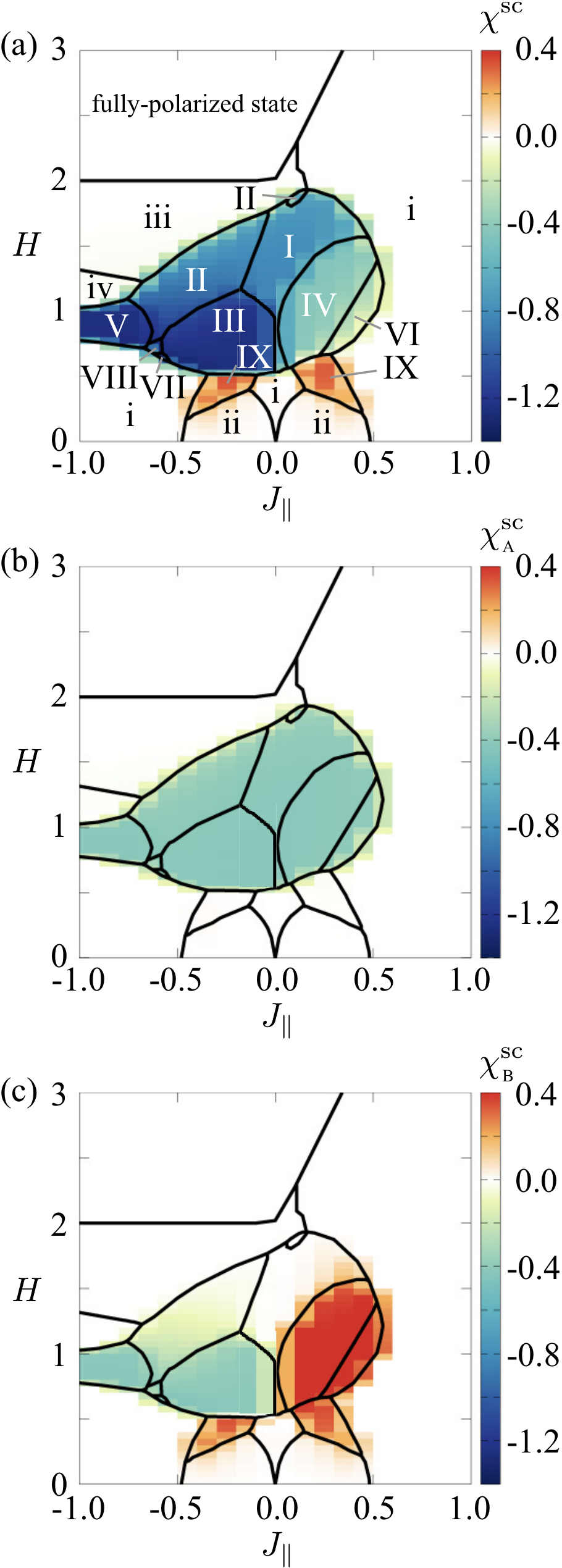} 
\caption{
\label{fig: PD}
(a) Phase diagram of the model in Eq.~(\ref{eq: Ham2}) obtained by the simulated annealing while changing $J_{\parallel}$ and $H$. 
The contour shows the total scalar chirality $\chi^{\rm sc}$. 
The layer-resolved scalar chiralities, $\chi^{\rm sc}_{\rm A}$ and $\chi^{\rm sc}_{\rm B}$, are plotted by the contour in (b) and (c), respectively, on the same phase diagram as (a).  
}
\end{center}
\end{figure}

\begin{table*}[htb!]
\centering
\caption{
Classification of the ordered phases in the model in Eq.~(\ref{eq: Ham2}) obtained by the simulated annealing. 
$n_{\rm sk}^{(\eta)}$ represents the skyrmion number for layer $\eta=$ A, B, and C, and $n_{\rm sk}^{\rm ave}$ represents the averaged skyrmion number. 
The other quantities are defined in Eqs.~(\ref{eq:Sq}) and (\ref{eq: chirality}). 
In the columns of $S^{xy}_{\eta }(\bm{Q}_{\nu})$ and $S^{z}_{\eta }(\bm{Q}_{\nu})$, $Q'$ stands for the different intensities of the $\bm{Q}_\nu$ components. 
\label{table: OP}}
\renewcommand{\arraystretch}{2}
\begin{tabular}{lccccccccccc}\hline \hline
name  & $n^{\rm (A)}_{\rm sk}$& $n^{\rm (B)}_{\rm sk}$& $n^{\rm (C)}_{\rm sk}$& $n^{\rm ave}_{\rm sk}$ &$\chi^{\rm sc}$ & $S^{xy}_{{\rm A}}(\bm{Q}_\nu)$ & $S^{z}_{{\rm A}}(\bm{Q}_\nu)$ & $S^{xy}_{{\rm B}}(\bm{Q}_\nu)$
 & $S^{z}_{{\rm B}}(\bm{Q}_\nu)$ & $S^{xy}_{{\rm C}}(\bm{Q}_\nu)$ & $S^{z}_{{\rm C}}(\bm{Q}_\nu)$ \\ \hline
Phase I &  $-1$ & $0$ & $-1$ & $\displaystyle -\frac{2}{3}$ & $\checkmark$ & $3Q'$& $3Q'$& $2Q$& $1Q$ & $3Q'$& $3Q'$\\ 
Phase II & $-1$ & $0$ & $-1$ & $\displaystyle -\frac{2}{3}$ & $\checkmark$ & $3Q$& $3Q$& $3Q$& $3Q$ & $3Q$& $3Q$\\ 
Phase III & $-1$ & $-1$ & $-1$ & $\displaystyle -1$ & $\checkmark$ & $3Q$& $3Q$& $3Q$& $3Q$ & $3Q$& $3Q$\\ 
Phase IV & $-1$ & $1$ & $-1$ & $\displaystyle -\frac{1}{3}$ & $\checkmark$ & $3Q'$& $3Q'$& $3Q'$& $3Q'$ & $3Q'$& $3Q'$\\ 
Phase V &  $-1$ & $-1$ & $-1$ & $\displaystyle -1$  & $\checkmark$ & $3Q$& $3Q$& $3Q$& $3Q$ & $3Q$& $3Q$\\ 
Phase VI &  $-1$ & $2$ & $-1$ & $\displaystyle 0$   & $\checkmark$ & $3Q'$& $3Q'$& $2Q$& $1Q$ & $3Q'$& $3Q'$\\
Phase VII & $-1$ & $-2$ & $-1$ & $\displaystyle -\frac{4}{3}$ & $\checkmark$ & $3Q$& $3Q$& $3Q$& $3Q$ & $3Q$& $3Q$\\
Phase VIII & $-1$ & $x$\footnote{$-2<x<0$} & $-1$ & $\displaystyle -\frac{2-x}{3}$ & $\checkmark$ & $3Q$& $3Q$& $3Q$& $3Q$ & $3Q$& $3Q$\\ 
Phase IX &  $0$ & $1$ & $0$ & $\displaystyle \frac{1}{3}$ & $\checkmark$ & $3Q'$& $1Q$\footnote{Negligebly small intensities are found at the other $\bm{Q}_\eta$ components.}& $3Q'$& $3Q'$ & $3Q'$& $1Q$\footnotemark[2] \\      \hline
Phase i & $0$ & $0$ & $0$ & $0$ & No & $1Q$& $1Q$& $1Q$& No & $1Q$& $1Q$ \\ 
Phase ii & $0$ & $0$ & $0$ & $0$ & $\checkmark$ & $3Q'$& $2Q'$\footnotemark[2]& $3Q'$& $3Q'$ & $3Q'$& $1Q$\footnotemark[2]\\ 
Phase iii & $0$ & $0$ & $0$ & $0$ & $\checkmark$ & $3Q'$& $3Q'$& $2Q$& $1Q$ & $3Q'$& $3Q'$\\ 
Phase iv & $0$ & $0$ & $0$ & $0$ & $\checkmark$ & $3Q'$& $3Q'$& $2Q'$& $1Q$ & $3Q'$& $3Q'$\\ 
\hline \hline
\end{tabular}
\end{table*}

We discuss the results obtained by the simulated annealing for the spin model in Eq.~(\ref{eq: Ham2}). 
Figure~\ref{fig: PD}(a) shows the low-temperature phase diagram while changing $J_{\parallel}$ and $H$. 
The phase boundaries are determined by changing $H$ by 0.0125 and $J_{\parallel}$ by 0.025 in their vicinity regions.
The color stands for the total scalar chirality $\chi^{\rm sc}$. 
As shown in the phase diagram in Fig.~\ref{fig: PD}(a), we find different thirteen phases with distinct spin and scalar chirality configurations in addition to the fully-polarized state, whose magnetic moments are along the $z$ direction. 
Among them, we find that nine out of thirteen phases possess a quantized skyrmion number for any of the layers. 
We label these nine phases by using the uppercase roman numerals as ``Phase I", ``Phase II", $\cdots$, ``Phase IX", while we denote the other four phases by using the lowercase roman numerals as ``Phase i", ``Phase ii", ``Phase iii", and ``Phase iv". 

The nine types of the SkX phases emerge under the external magnetic field from the ferromagnetic interlayer exchange interaction to the antiferromagnetic one. 
The region where the SkXs are stabilized is asymmetric regarding the sign of $J_{\parallel}$; the critical value of $|J_{\parallel}|$ for the ferromagnetic exchange interaction is larger than that for the antiferromagnetic one. 
Although the appearance of the SkX phases in the trilayer system is common to that in the bilayer system, only the single SkX phase is realized in the bilayer system~\cite{Hayami_PhysRevB.105.014408}. 
Thus, the present trilayer system consisting of the inversion-symmetric layer (layer B) and two inversion-asymmetric layers (layers A and C) enables us to engineer and design multiple SkX phases. 

Especially, we find that the scalar chirality for layer B is sensitive against the change of $J_{\parallel}$ and $H$ compared to those for layers A and C. 
We show the contour plots of the sublattice-dependent scalar chiralities $\chi^{\rm sc}_{\rm A}$ and $\chi^{\rm sc}_{\rm B}$ in Figs.~\ref{fig: PD}(b) and \ref{fig: PD}(c), respectively. 
The behavior of $\chi^{\rm sc}_{\rm C}$ is almost the same as that of $\chi^{\rm sc}_{\rm A}$. 
The value of $\chi^{\rm sc}_{\rm B}$ takes both positive and negative values in Fig.~\ref{fig: PD}(c), while that of $\chi^{\rm sc}_{\rm A}$ takes negative values with almost constant magnitude in Fig.~\ref{fig: PD}(b). 
The latter behavior regarding layer A is roughly consistent with that in the bilayer system where only the single SkX phase appears. 
Indeed, the skyrmion number in the colored region from Phase I to Phase VIII in Fig.~\ref{fig: PD}(b) remains the same. 
On the other hand, there are multiple skyrmion numbers in the colored region for layer B, as discussed below.   
The obtained phase diagram indicates that the introduction of the inversion-symmetric middle layer (layer B) is a source of multiple SkX phases. 

The four phases denoted as Phase i to Phase iv are characterized by single-$Q$ or triple-$Q$ states with no skyrmion number. 
Among them, three phases except for Phase i take finite values of $\chi^{\rm sc}$, although their magnitudes are much smaller than those in the SkX phases; for example, see Figs.~\ref{Fig: Mag1} and \ref{Fig: Mag2}. 

In what follows, we describe details of the obtained phases one by one in Sec.~\ref{sec: Details of magnetic phases}. 
We list the skyrmion number $(n_{\rm sk}^{\rm (A)}, n_{\rm sk}^{\rm (B)}, n_{\rm sk}^{\rm (C)}, n_{\rm sk}^{\rm ave})$, scalar chirality $\chi^{\rm sc}$, and $xy$ and $z$ components of the spin structure factor $[S^{xy}_\eta(\bm{Q}_\nu), S^{z}_\eta(\bm{Q}_\nu)]$ for $\eta=$ A, B, and C in each obtained phase in Table~\ref{table: OP} for reference. 
In addition, the real-space spin and scalar chirality configurations in each phase are shown in Figs.~\ref{Fig:spin1}-\ref{Fig:spin7}, and their corresponding spin structure factors are shown in Figs.~\ref{Fig:sq1} and \ref{Fig:sq2}. 
We also discuss the $H$ dependences of the magnetization and scalar chirality for several $J_{\parallel}$ in Sec.~\ref{sec: Magnetic-field dependence}. 

\subsection{Details of magnetic phases}
\label{sec: Details of magnetic phases}

\begin{figure*}[htb!]
\begin{center}
\includegraphics[width=1.0 \hsize]{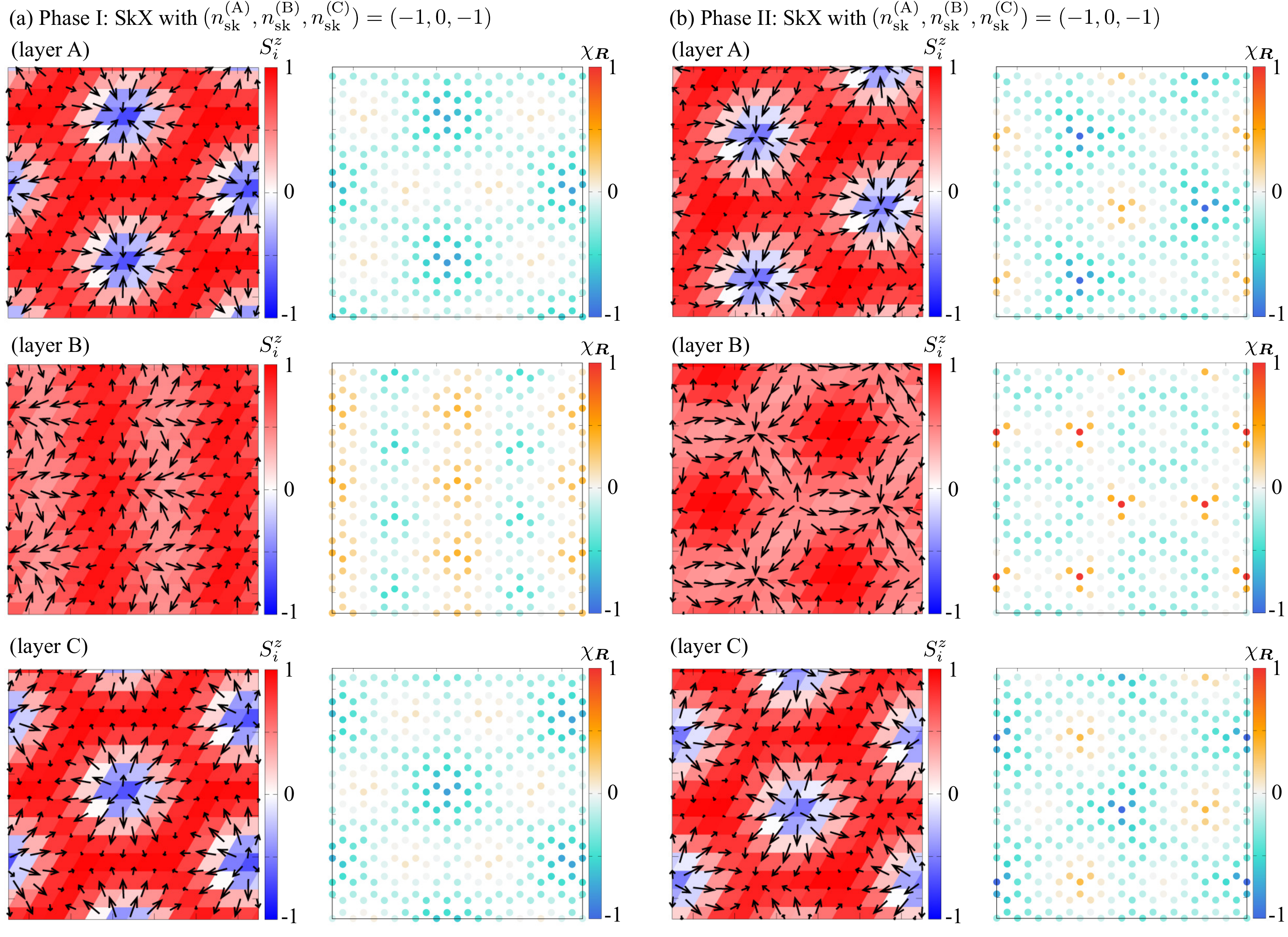} 
\caption{
\label{Fig:spin1}
Real-space spin (left panel) and scalar chiraity (right panel) configurations of (a) Phase I [SkX with $(n_{\rm sk}^{\rm (A)}, n_{\rm sk}^{\rm (B)}, n_{\rm sk}^{\rm (C)})=(-1,0,-1)$] at $J_{\parallel}=0.1$ and $H=1.5$ and (b) Phase II [SkX with $(n_{\rm sk}^{\rm (A)}, n_{\rm sk}^{\rm (B)}, n_{\rm sk}^{\rm (C)})=(-1,0,-1)$] at $J_{\parallel}=-0.5$ and $H=1.2$ on layer A (upper panel), layer B (middle panel), and layer C (lower panel).  
In the left panels, the arrows represent the $xy$ components of the spin moment and the color shows the $z$ component. 
}
\end{center}
\end{figure*}

\begin{figure*}[htb!]
\begin{center}
\includegraphics[width=0.97 \hsize]{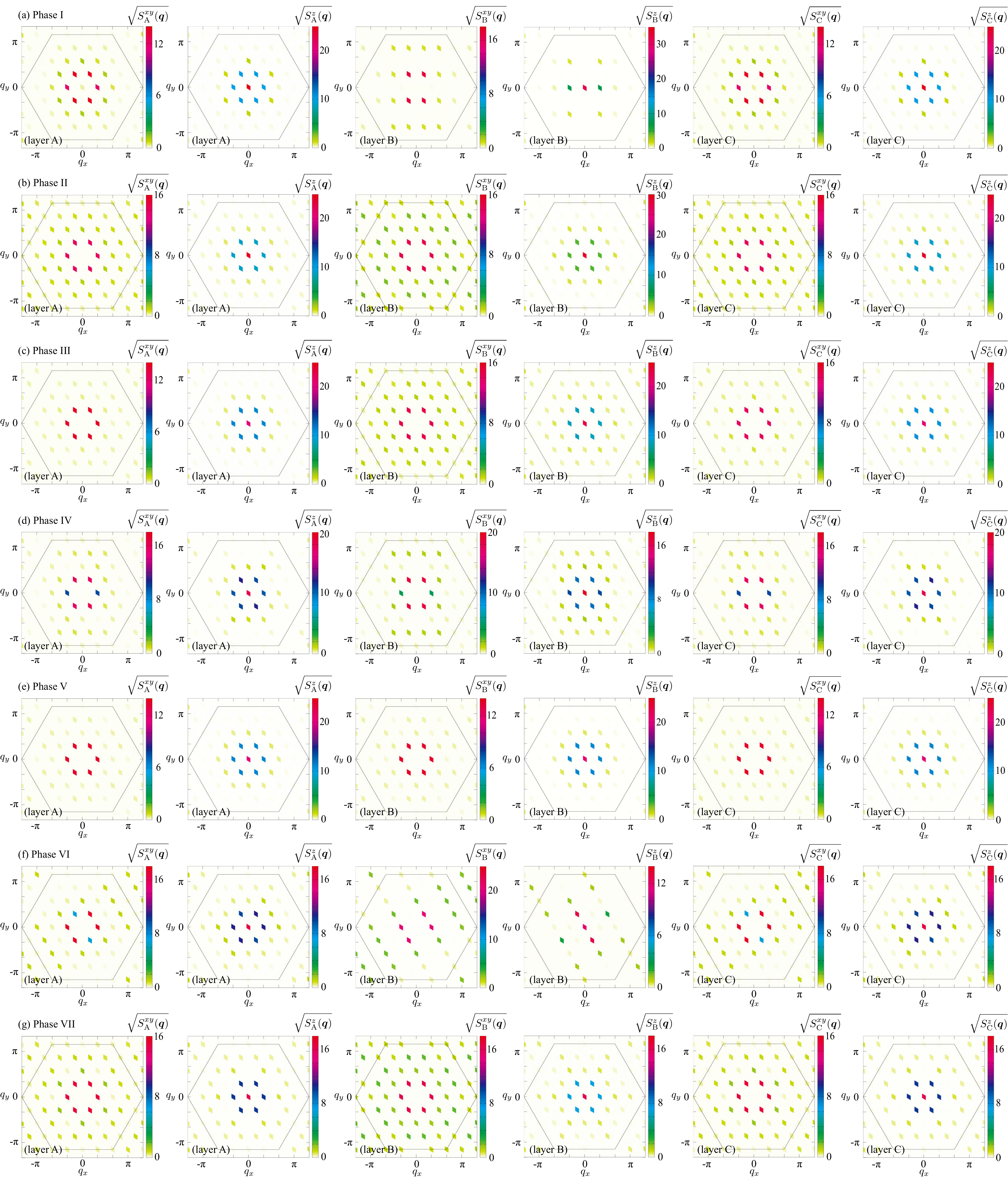} 
\caption{
\label{Fig:sq1}
(Left and second left) The square root of the $xy$ and $z$ components of the spin structure factor for layer A, respectively, in (a) Phase I at $J_{\parallel}=0.1$ and $H=1.5$, (b) Phase II at $J_{\parallel}=-0.5$ and $H=1.2$, (c) Phase III at $J_{\parallel}=-0.2$ and $H=1$, (d) Phase IV at $J_{\parallel}=0.2$ and $H=1$, (e) Phase V at $J_{\parallel}=-1$ and $H=0.9$, (f) Phase VI at $J_{\parallel}=0.4$ and $H=1$, and (g) phase VII state at $J_{\parallel}=-0.6$ and $H=0.65$.
Black hexagons represent the first Brillouin zone.   
The middle and right two panels represent the data for layer B and layer C corresponding to the left two ones, respectively. 
}
\end{center}
\end{figure*}

\paragraph{Phase I: SkX with $(n_{\rm sk}^{\rm (A)}, n_{\rm sk}^{\rm (B)}, n_{\rm sk}^{\rm (C)})=(-1,0,-1)$.}
This state is stabilized for small $|J_{\parallel}|$ and intermediate $H$ in Fig.~\ref{fig: PD}(a), which extends to the positive $J_{\parallel}$ region rather than the negative one. 
As shown in the real-space spin configuration in the left panel of Fig.~\ref{Fig:spin1}(a), this state is characterized by a coexisting state of the N\'eel SkX for layers A and C and the triple-$Q$ state for layer B; the skyrmion core denoted at $S_i^z \simeq -1$ forms the triangular lattice for layers A and C, although their positions are different from each other. 
The helicity of the SkXs for layers A and C is opposite owing to the opposite sign of the DM interaction. 
According to the SkX spin textures, the scalar chirality seems to be distributed in an almost threefold-symmetric way, which results in the skyrmion number of $-1$. 
The formation of the SkX is also found in the spin structure factor in Fig.~\ref{Fig:sq1}(a); both $xy$ and $z$ components exhibit almost triple-$Q$ peak structures. 
It is noted that the intensities at the triple-$Q$ wave vectors are slightly different due to the coupling to layer B, whose spin texture breaks the threefold symmetry of the triangular lattice, as shown in the middle panel of Fig.~\ref{Fig:spin1}(a).
We use the symbol $Q'$ to represent the different intensities of the $\bm{Q}_\nu$ component in Table~\ref{table: OP}. 
The spin configuration for layer B is mainly characterized by the double-$Q$ peaks in the $xy$ component and the single-$Q$ peak in the $z$ component, as shown in Fig.~\ref{Fig:sq1}(a). 
The scalar chirality for layer B behaves like a chirality density wave along the $\bm{Q}_1$ direction. 
Although there is no skyrmion number for layer B, the scalar chirality takes a nonzero positive value. 
The nonzero scalar chirality for layer B is owing to the spiral modulation via the interlayer coupling in the multi-layer system. 
Indeed, it vanishes when $J_{\parallel}=0$. 

In the vicinity of $J_{\parallel}=0$, one of the double-$Q$ peaks in the $xy$ spin component and the single-$Q$ peak in the $z$ spin component for layer B are suppressed while decreasing $|J_{\parallel}|$, and then, the spin configuration turns into the single-$Q$ conical spiral state for $J_{\parallel}=0$, whose spiral plane lies in the $xy$ plane. 
Accordingly, the intensities of the spin structure factor at triple-$Q$ wave vectors for layers A and C are equivalent, which indicates the recovery of the threefold rotational symmetry.

\paragraph{Phase II: SkX with $(n_{\rm sk}^{\rm (A)}, n_{\rm sk}^{\rm (B)}, n_{\rm sk}^{\rm (C)})=(-1,0,-1)$.}
Phase II mainly emerges for negative $J_{\parallel}$ and intermediate $H$, which is obtained next to Phase I upon decreasing $J_{\parallel}$. 
Although this phase also appears in the narrow region for positive $J_{\parallel}$ sandwiched by Phase I and Phase iii, we here focus on the region for $J_{\parallel}<0$. 
This state exhibits similar spin configurations for layers A and C to those in Phase I, as shown in Figs.~\ref{Fig:spin1}(a) and \ref{Fig:spin1}(b). 
The main difference between Phase I and Phase II is found in the spin configuration for layer B: 
The former is characterized by the anisotropic triple-$Q$ spin configuration, while the latter is by the isotropic one satisfying the threefold rotational symmetry. 
As a result, the spin texture in Phase II is invariant under the threefold rotation. 
Such a feature is also found in the spin structure factor in Fig.~\ref{Fig:sq1}(b). 
Both $xy$ and $z$ components of spin exhibit the triple-$Q$ peak structures with the same intensities. 
Meanwhile, the skyrmion number in each layer in Phase II is the same as that in Phase I, i.e., $(n_{\rm sk}^{\rm (A)}, n_{\rm sk}^{\rm (B)}, n_{\rm sk}^{\rm (C)})=(-1,0,-1)$. 
In other words, the spin configuration for layer B exhibits no skyrmion number, although the scalar chirality takes a negative value. 

By closely looking at the real-space spin configuration in each layer in Fig.~\ref{Fig:spin1}(b), one finds that the inplane spin structure at the same $(x, y)$ position is almost the same as each other. 
Thus, the different spin configurations in different layers arise from the $z$-spin component, which is attributed to the phase degree of freedom in the multiple-$Q$ superposition~\cite{yambe2021skyrmion, hayami2021phase, Hayami_PhysRevResearch.3.043158,shimizu2022phase}.
The difference in the spin textures between layers A and C is the relative phase degree of freedom among the constituent spiral waves at different wave vectors, while that between layers A and B is the phase degree of freedom in terms of $xy$ and $z$ spins~\cite{yambe2021skyrmion}. 
In addition, the tendency that aligns the $xy$-spin component rather than the $z$-spin one between adjacent layers is energetically understood from the large amplitude of the $xy$ spin component compared to that of the $z$ spin, as shown in Fig.~\ref{Fig:sq1}(b).

\begin{figure*}[htb!]
\begin{center}
\includegraphics[width=1.0 \hsize]{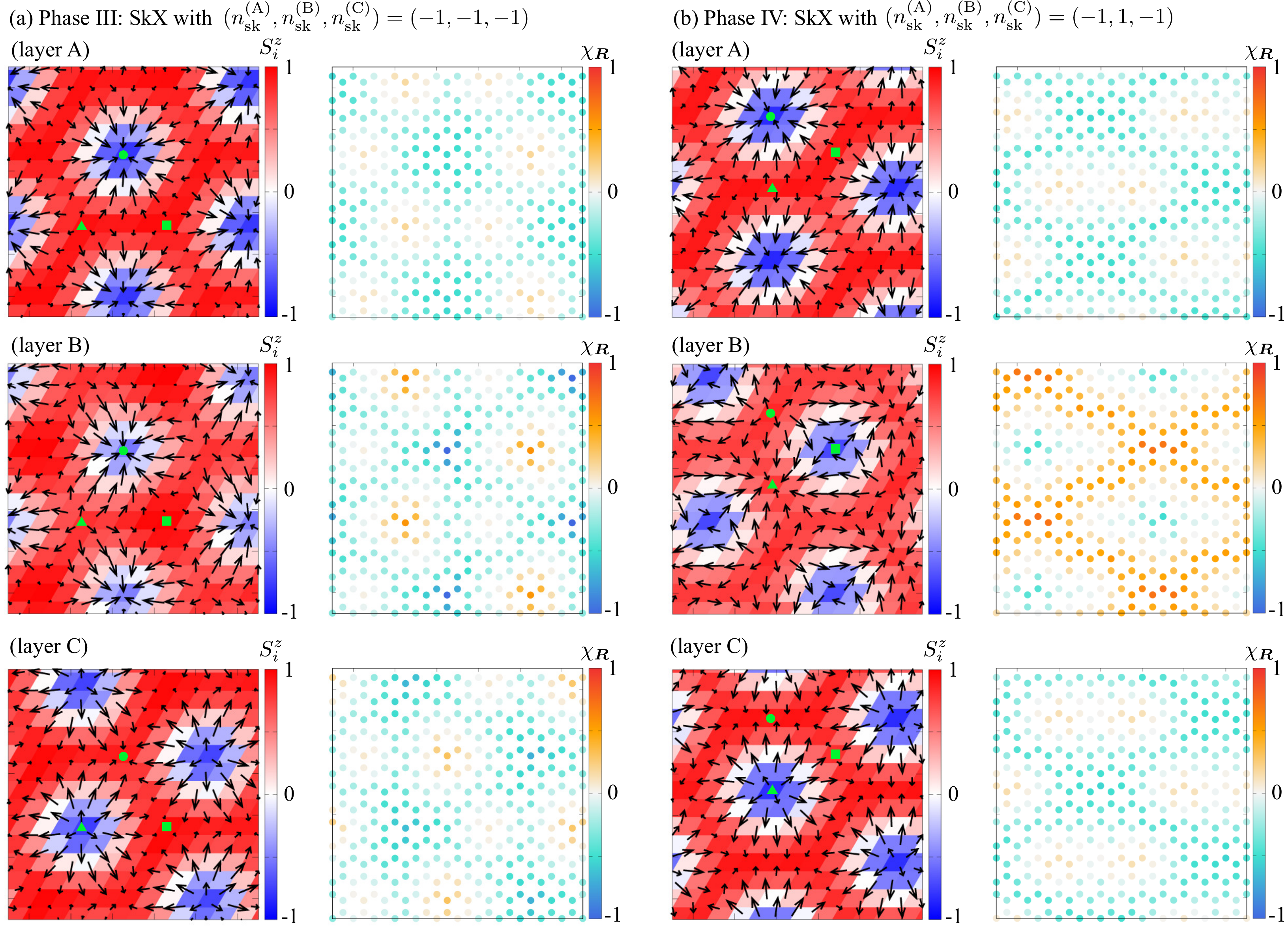} 
\caption{
\label{Fig:spin2}
Real-space spin (left panel) and scalar chiraity (right panel) configurations of (a) Phase III [SkX with $(n_{\rm sk}^{\rm (A)}, n_{\rm sk}^{\rm (B)}, n_{\rm sk}^{\rm (C)})=(-1,-1,-1)$] at $J_{\parallel}=-0.2$ and $H=1$ and (b) Phase IV [SkX with $(n_{\rm sk}^{\rm (A)}, n_{\rm sk}^{\rm (B)}, n_{\rm sk}^{\rm (C)})=(-1,1,-1)$] at $J_{\parallel}=0.2$ and $H=1$ on layer A (upper panel), layer B (middle panel), and layer C (lower panel).  
In the left panels, the arrows represent the $xy$ components of the spin moment and the color shows the $z$ component. 
The three types of vortex cores are represented by green circles, triangles, and squares. 
}
\end{center}
\end{figure*}

 \paragraph{Phase III: SkX with $(n_{\rm sk}^{\rm (A)}, n_{\rm sk}^{\rm (B)}, n_{\rm sk}^{\rm (C)})=(-1,-1,-1)$.}
Phase III appears for the ferromagnetic interlayer exchange interaction $J_{\parallel}<0$ upon decreasing $H$ from Phase I and Phase II, as shown in Fig.~\ref{fig: PD}(a). 
In contrast to Phase I and Phase II, the spin texture for layer B turns into the SkX one, as shown in Fig.~\ref{Fig:spin2}(a). 
The skyrmion core position and the helicity for layer B are the same as those for layer A or layer C depending on initial random spin configurations. 
Thus, the skyrmion number in Phase III is given by $(n_{\rm sk}^{\rm (A)}, n_{\rm sk}^{\rm (B)}, n_{\rm sk}^{\rm (C)})=(-1,-1,-1)$. 
The spin structure factor exhibits the triple-$Q$ peak structure in both $xy$ and $z$ components with equal intensity, as shown in Fig.~\ref{Fig:sq1}(c), although their intensities are different for different layers. 
In the scalar chirality sector, the real-space distribution of $\chi_{\bm{R}}$ keeps the threefold rotational symmetry. 
The magnitude of $\chi_{\eta}^{\rm sc}$ is slightly different from each other, as shown in Fig.~\ref{Fig: Mag1}.

The appearance of the spin configuration in Phase III is owing to the frustration that arises from the competition between the ferromagnetic interlayer exchange interaction and the layer-dependent DM interaction: The former favors the same direction of the spin moments, whereas the latter favors the opposite spin direction in the $xy$ component owing to the different helicity for layers A and C. 
The obtained spin configuration in Fig.~\ref{Fig:spin2}(a) is a consequence of the optimization of the energies by both interactions as follows. 
When the spin configuration is regarded as the periodic alignment of three types of vortices denoted as green circles, triangles, and squares in Fig.~\ref{Fig:spin2}(a), the skyrmion cores are located around the circles for layers A and B and around the triangles for layer C. 
With respect to the interlayer exchange coupling, the $z$-spin component around the cores denoted by the circles and triangles does not lead to an energy gain because the number of parallel spin pairs is the same as that of anti-parallel spin pairs. 
Instead of that, there is an energy gain by $J_{\parallel}$ for the $xy$-spin component; the different skyrmion core positions for layers A and C lead to an energy gain by the DM interaction. 
In addition, there is an energy gain by $J_{\parallel}$ for both the $xy$- and $z$-spin components around the cores denoted by the squares.

\paragraph{Phase IV: SkX with $(n_{\rm sk}^{\rm (A)}, n_{\rm sk}^{\rm (B)}, n_{\rm sk}^{\rm (C)})=(-1,1,-1)$.}
This state appears only for the antiferromagnetic exchange interaction $J_{\parallel}>0$ while decreasing $H$ from Phase I. 
Notably, the spin configuration in Phase IV consists of the SkX for layers A and C and the anti-SkX for layer B, which take the opposite sign of the skyrmion number. 
Indeed, the negative (positive) contribution of the local scalar chirality $\chi_{\bm{R}}$ is dominant for layers A and C (layer B), as shown in the right panel of Fig.~\ref{Fig:spin2}(b). 
Since the anti-SkX in the middle layer breaks the threefold rotational symmetry of the triangular lattice, the other two layers do not also possess the threefold axis via the coupling to layer B, and thus, the spin structure factor shows the anisotropic triple-$Q$ peak structures in Fig.~\ref{Fig:sq1}(d). 
The intensities of the spin structure factor for layers A and C are equal, which are different from those for layer B. 
Accordingly, the magnetization and the scalar chirality for layers A and C take different values from those for layer B, as shown in Fig.~\ref{Fig: Mag2}. 
 
The emergence of the anti-SkX for layer B in Phase IV is understood by considering the energy gain around the vortex cores. 
Similar to Phase III, we consider three types of vortices, but we focus on the different core positions, as compared to Figs.~\ref{Fig:spin2}(a) and \ref{Fig:spin2}(b). 
For layer A, the cores denoted by the circle, triangle, and square correspond to the skyrmion core, vortex core, and anti-vortex core, respectively, where the vortex and anti-vortex cores have the opposite winding numbers from each other. 
For layer C, the positions of the skyrmion and vortex cores for layer A are exchanged so as to gain energy by the DM interaction. 
In these situations, the spin texture for layer B is determined so as to gain the energy by the antiferromagnetic interlayer exchange interaction; the spins around two out of three cores show the anti-parallel alignment; specifically, the spins around the circle (triangle) and square cores point along the opposite directions between layers A and B (B and C)

It is noted that the anti-SkX spin texture with the positive skyrmion number that appears in layer B is rare in the triangular-lattice system. 
This is because such a spin texture breaks the threefold rotational symmetry of the triangular lattice in the presence of the anisotropic exchange interaction so that the spin and orbit (lattice) degrees of freedom are entangled, which usually results in a higher energy than the Bloch or N\`eel SkX with the threefold axis. 
In the present situation, the anti-SkX is brought about by effective mean fields breaking threefold rotational symmetry via the antiferromagnetic interlayer exchange coupling under the SkXs with different helicities for layers A and C. 
Thus, the synergy between the opposite sign of the DM interaction and the antiferromagnetic interlayer exchange interaction plays a significant role in realizing the anti-SkX spin texture for layer B. 
 
\begin{figure*}[htb!]
\begin{center}
\includegraphics[width=1.0 \hsize]{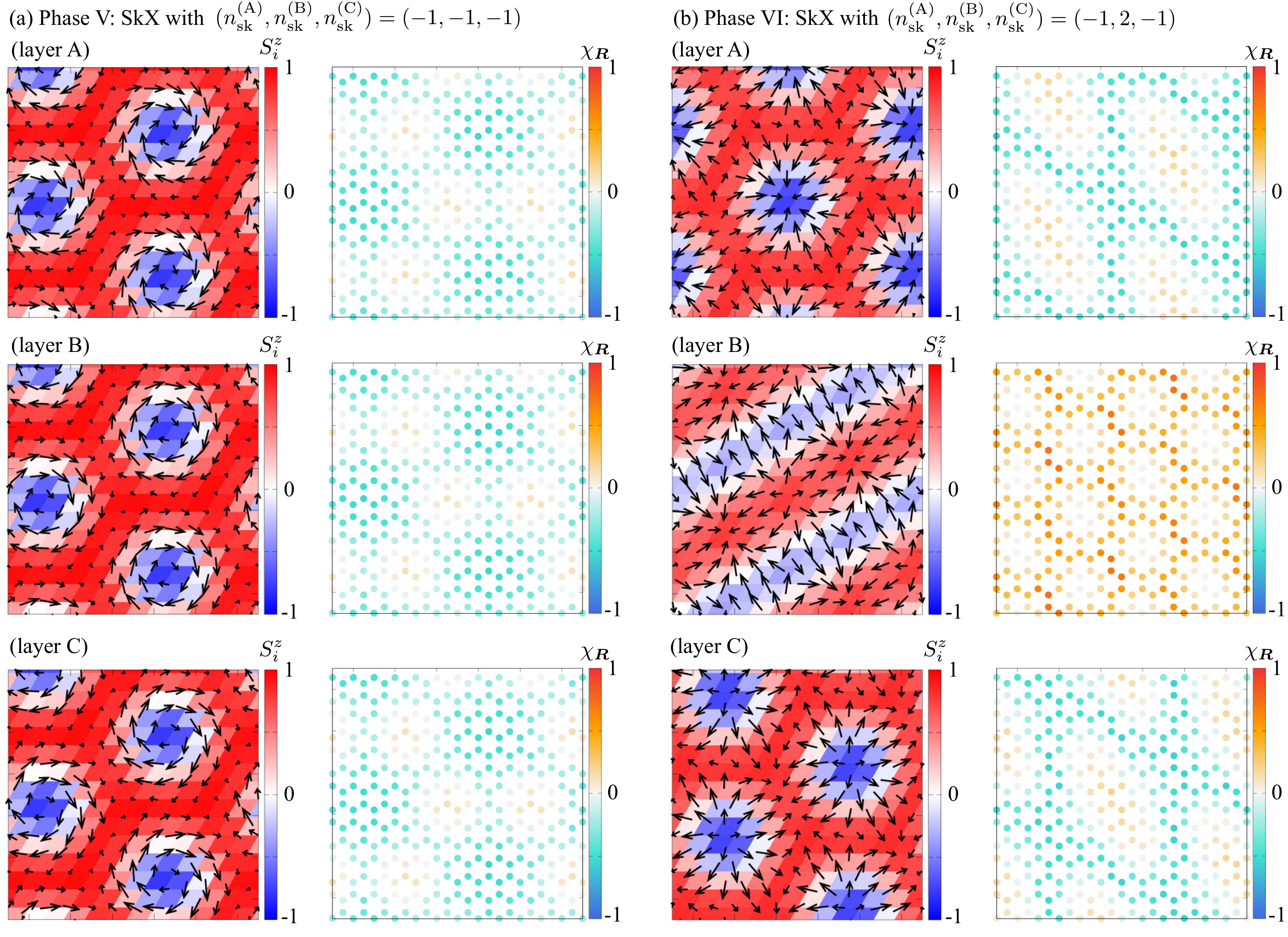} 
\caption{
\label{Fig:spin3}
Real-space spin (left panel) and scalar chiraity (right panel) configurations of (a) Phase V [SkX with $(n_{\rm sk}^{\rm (A)}, n_{\rm sk}^{\rm (B)}, n_{\rm sk}^{\rm (C)})=(-1,-1,-1)$] at $J_{\parallel}=-1$ and $H=0.9$ and (b) Phase VI [SkX with $(n_{\rm sk}^{\rm (A)}, n_{\rm sk}^{\rm (B)}, n_{\rm sk}^{\rm (C)})=(-1,2,-1)$] at $J_{\parallel}=0.4$ and $H=1$ on layer A (upper panel), layer B (middle panel), and layer C (lower panel).  
In the left panels, the arrows represent the $xy$ components of the spin moment and the color shows the $z$ component. 
}
\end{center}
\end{figure*}

\begin{figure*}[htb!]
\begin{center}
\includegraphics[width=1.0 \hsize]{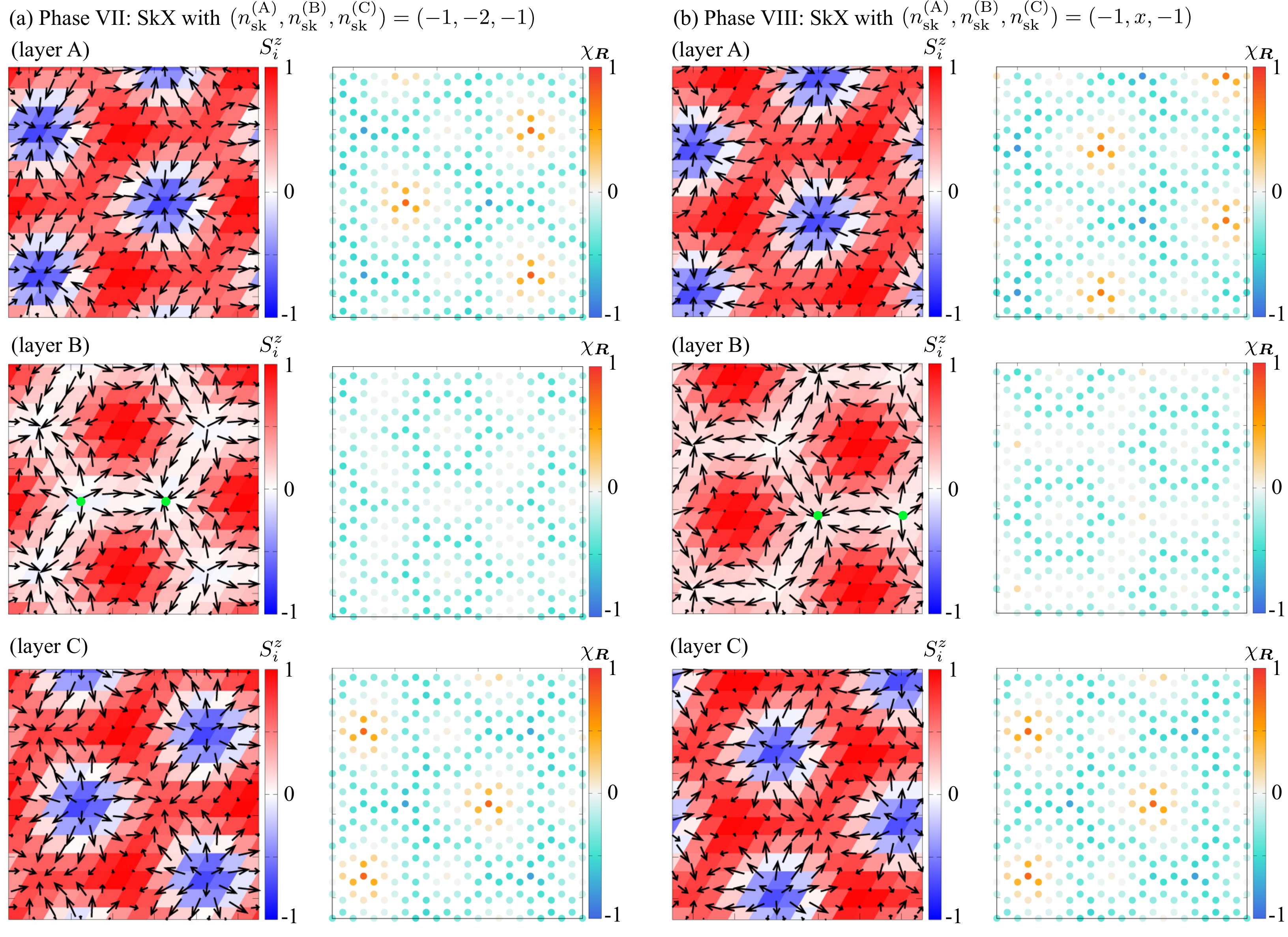} 
\caption{
\label{Fig:spin4}
Real-space spin (left panel) and scalar chiraity (right panel) configurations of (a) Phase VII [SkX with $(n_{\rm sk}^{\rm (A)}, n_{\rm sk}^{\rm (B)}, n_{\rm sk}^{\rm (C)})=(-1,-2,-1)$] at $J_{\parallel}=-0.6$ and $H=0.65$ and (b) Phase VIII [SkX with $(n_{\rm sk}^{\rm (A)}, n_{\rm sk}^{\rm (B)}, n_{\rm sk}^{\rm (C)})=(-1,x,-1)$] at $J_{\parallel}=-0.6$ and $H=0.75$ on layer A (upper panel), layer B (middle panel), and layer C (lower panel).  
In the left panels, the arrows represent the $xy$ components of the spin moment and the color shows the $z$ component. 
The green circles stand for the core positions. 
}
\end{center}
\end{figure*}

\paragraph{Phase V: SkX with $(n_{\rm sk}^{\rm (A)}, n_{\rm sk}^{\rm (B)}, n_{\rm sk}^{\rm (C)})=(-1,-1,-1)$.}
Phase V appears for the large ferromagnetic interlayer exchange interaction and intermediate $H$, which is obtained when decreasing $J_{\parallel}$ from Phase II [Fig.~\ref{fig: PD}(a)]. 
This phase consists of three SkX layers with the skyrmion number of $-1$, as shown in Fig.~\ref{Fig:spin3}(a). 
In contrast to the other SkX phases, the core positions and the helicity around the skyrmion core are almost the same for the three layers. 
Especially, the helicity around the skyrmion core is different from the N\'eel SkX; the spins around the core are twisted from the N\'eel type to the Bloch type. 
Such twisted spin arrangement around the core is owing to the surface effect for the upper and lower layers, which has been discussed in the context of the twisted surface SkX~\cite{PhysRevB.87.094424, PhysRevLett.120.227202,bo2021tailoring}. 
This spin state is realized as a result of the energy gain by the ferromagnetic interlayer exchange interaction rather than the DM interaction. 
The spin structure factor exhibits the triple-$Q$ peak structure in both $xy$ and $z$ spin components, which is similar to that in Phase III, as shown in Fig.~\ref{Fig:sq1}(e).  
The magnetization and scalar chirality in each layer are almost the same as each other, as shown in Fig.~\ref{Fig: Mag1}. 

\paragraph{Phase VI: SkX with $(n_{\rm sk}^{\rm (A)}, n_{\rm sk}^{\rm (B)}, n_{\rm sk}^{\rm (C)})=(-1,2,-1)$.}
This phase appears next to Phase IV upon increasing $J_{\parallel}$ or decreasing $H$. 
The real-space spin configurations for layers A and C are similar to those in Phase IV, as compared in Figs.~\ref{Fig:spin2}(b) and \ref{Fig:spin3}(b); the SkXs form the distorted triangular lattice, whose cores are located at different positions for layers A and C so that the threefold rotational symmetry is broken. 
This indicates that the intensities in the spin structure factor are different at $\bm{Q}_\nu$, as shown in Fig.~\ref{Fig:sq1}(f). 
The difference from Phase IV is found in the real-space spin and scalar chirality configurations for layer B, as shown in the middle panel in Fig.~\ref{Fig:spin3}(b). 
In this phase, the $xy$ component of spins shows the double-$Q$ structure with equal intensity at $\bm{Q}_1$ and $\bm{Q}_3$, while the $z$ component of spins shows the single-$Q$ structure at $\bm{Q}_2$, as shown in Fig.~\ref{Fig:sq1}(f). 
Although such a peak structure in the spin structure factor is similar to that for layer B in Phase I (see also Table~\ref{table: OP}), this spin configuration takes a high skyrmion number of two per magnetic unit cell, i.e., $n_{\rm sk}^{\rm (B)}=2$; almost all the regions exhibit the positive scalar chirality in the middle-right panel of Fig.~\ref{Fig:spin3}(b). 
The real-space spin ansatz for layer B is rougly given by 
\begin{align}
\bm{S}_i \propto [\cos (\bm{Q}_1\cdot \bm{r}_i),\cos (\bm{Q}_3\cdot \bm{r}_i),a^z \cos (\bm{Q}_2\cdot \bm{r}_i)],  
\end{align}
where the coefficient $a^z$ depends on the model parameters.
It is noted that this spin configuration has also been obtained in the Kondo lattice model with the single-ion anisotropy~\cite{Hayami_PhysRevB.99.094420} and its effective spin model~\cite{hayami2020multiple}, where the multiple-spin interactions become important. 
On the other hand, the present spin texture results from the interplay between the layer-dependent DM interaction and the interlayer antiferromagnetic exchange coupling within the bilinear spin interactions. 

As the skyrmion number for layer B is twice that for layers A and C and their sign is opposite, the averaged skyrmion number is zero, $n_{\rm sk}^{\rm ave}=0$. 
This indicates that the Hall conductivity is not quantized and vanishes in the case of the insulators. 
In this context, this state is regarded as an antiferromagnetic SkX, which has been investigated in the square, triangular, and honeycomb magnets~\cite{Rosales_PhysRevB.92.214439,zhang2016antiferromagnetic,Gobel_PhysRevB.96.060406,Diaz_hysRevLett.122.187203,Kravchuk_PhysRevB.99.184429,Tome_PhysRevB.103.L020403,mukherjee2021antiferromagnetic}. 
However, the present antiferromagnetic SkX (Phase VI) consists of different skyrmion numbers $-1$ and $2$, which is different from the previous findings.  
Thus, qualitatively different transport phenomena, such as the topological spin Hall effect, can be expected in Phase VI.

\begin{figure*}[htb!]
\begin{center}
\includegraphics[width=1.0 \hsize]{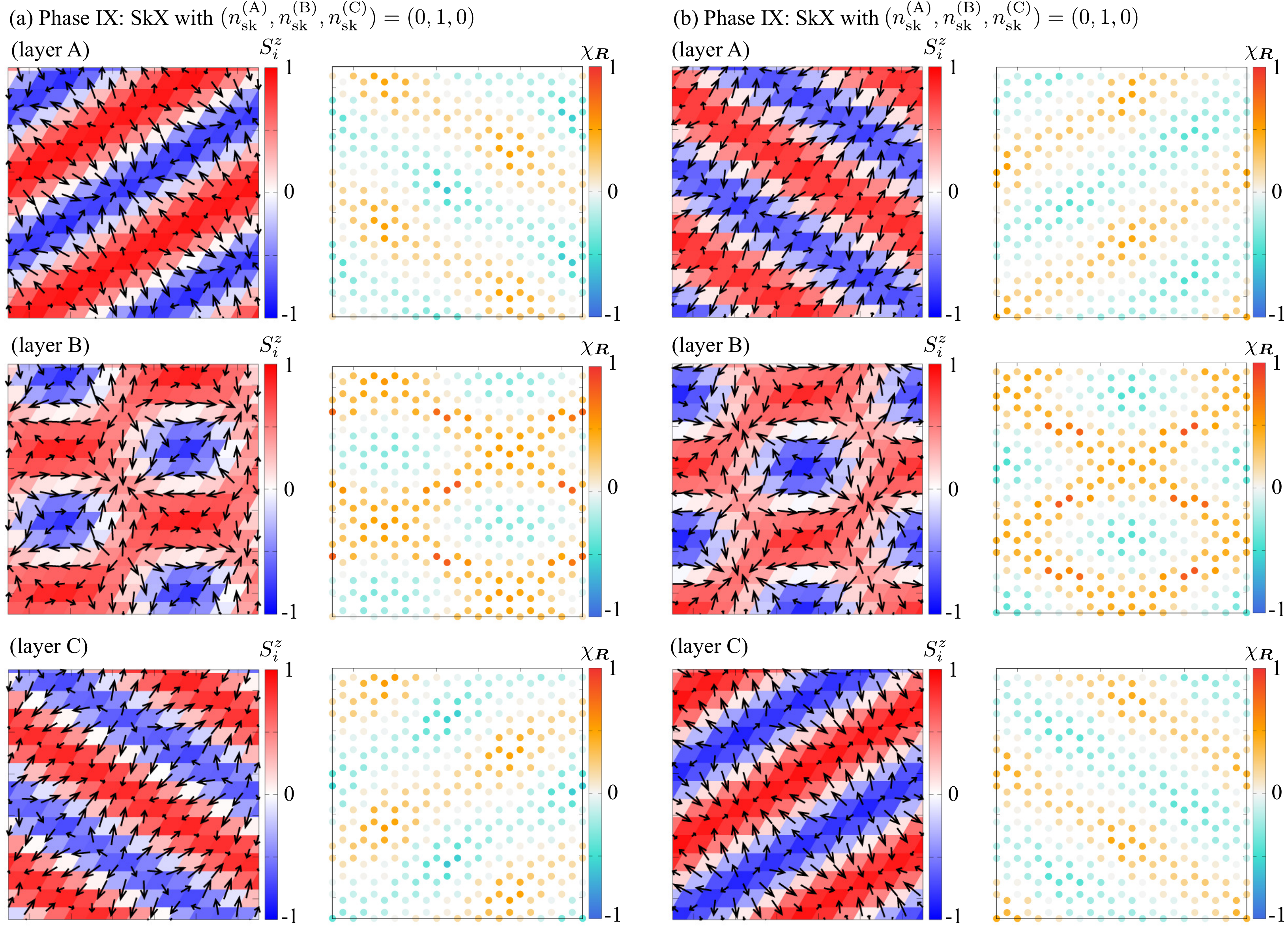} 
\caption{
\label{Fig:spin5}
Real-space spin (left panel) and scalar chiraity (right panel) configurations of (a) Phase IX [SkX with $(n_{\rm sk}^{\rm (A)}, n_{\rm sk}^{\rm (B)}, n_{\rm sk}^{\rm (C)})=(0,1,0)$] at $J_{\parallel}=-0.3$ and $H=0.4$ and (b) Phase IX [SkX with $(n_{\rm sk}^{\rm (A)}, n_{\rm sk}^{\rm (B)}, n_{\rm sk}^{\rm (C)})=(0,1,0)$] at $J_{\parallel}=0.3$ and $H=0.4$ on layer A (upper panel), layer B (middle panel), and layer C (lower panel).  
In the left panels, the arrows represent the $xy$ components of the spin moment and the color shows the $z$ component. 
}
\end{center}
\end{figure*}

\paragraph{Phase VII: SkX with $(n_{\rm sk}^{\rm (A)}, n_{\rm sk}^{\rm (B)}, n_{\rm sk}^{\rm (C)})=(-1,-2,-1)$.}
This phase emerges in the narrow region close to Phase III for $J_{\parallel}<0$, as shown in Fig.~\ref{fig: PD}(a). 
The real-space spin configuration in this phase [Fig.~\ref{Fig:spin4}(a)] is similar to that in Phase II [Fig.~\ref{Fig:spin1}(b)]. 
The spin structure factor is also similar to each other, as shown in Figs.~\ref{Fig:sq1}(b) and \ref{Fig:sq1}(g). 
The slight difference between them appears in the spin configuration for layer B in the middle panel of Fig.~\ref{Fig:spin4}(a). 
By closely looking at the spins around the vortex core denoted by the circles, one finds that the sign of the $z$-spin component becomes negative. 
This indicates that the sign of the scalar chirality around the core is reversed, as found from the comparison in the middle-right panel of Figs.~\ref{Fig:spin4}(a) and \ref{Fig:spin1}(b).  
Such a difference gives rise to the skyrmion number of $-2$ for layer B in Phase VII.
As the $z$ spin component around the core is small, its sign is reversed with a small change of $H$. 
This is why Phase VII is stabilized only in a narrow region compared to the other phases. 

\begin{figure*}[htb!]
\begin{center}
\includegraphics[width=0.97 \hsize]{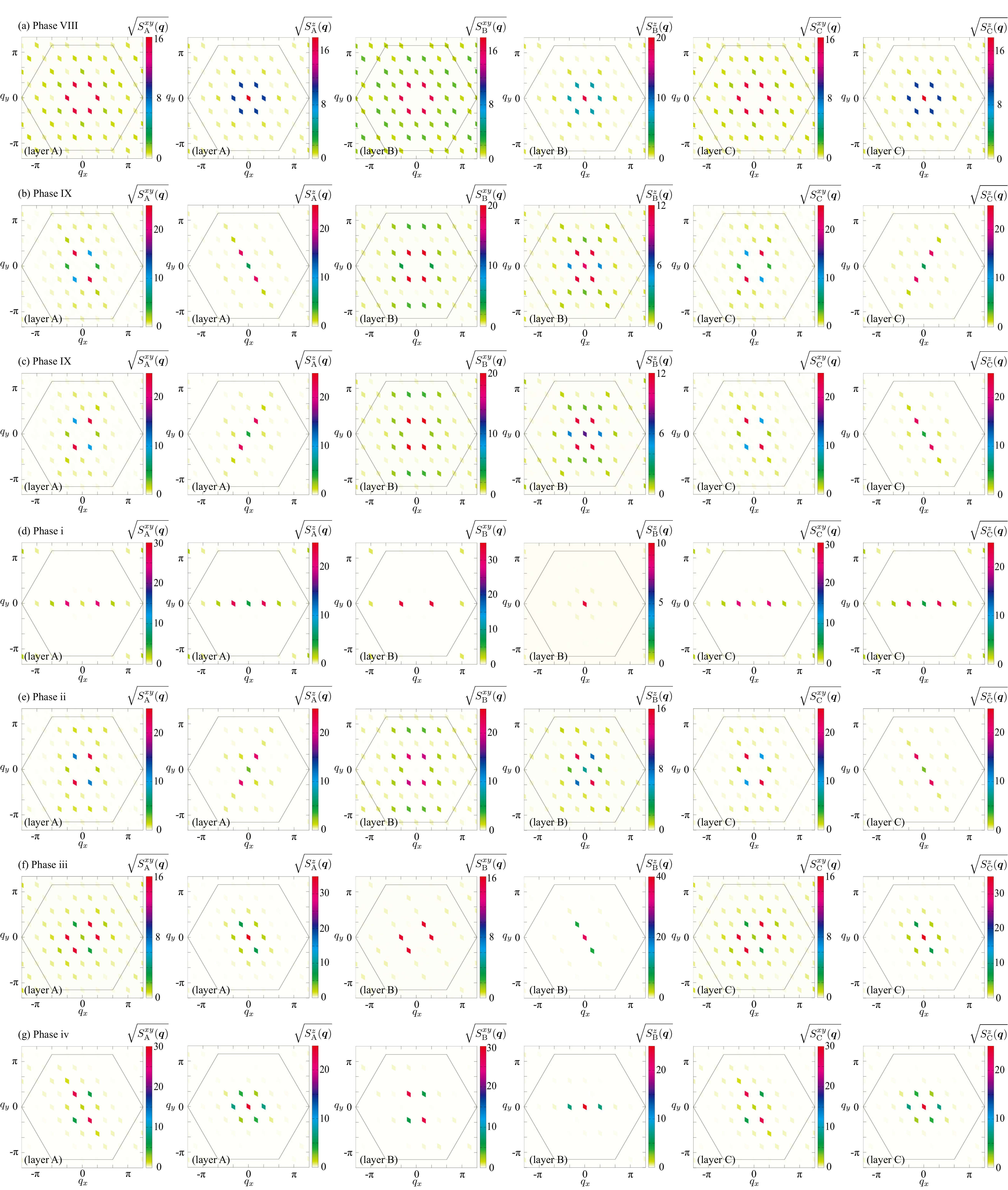} 
\caption{
\label{Fig:sq2}
(Left and second left) The square root of the $xy$ and $z$ components of the spin structure factor for layer A, respectively, in (a) Phase VIII at $J_{\parallel}=-0.6$ and $H=0.75$, (b) Phase IX at $J_{\parallel}=-0.3$ and $H=0.4$, (c) Phase IX at $J_{\parallel}=0.3$ and $H=0.4$, (d) Phase i at $J_{\parallel}=-0.1$ and $H=0.4$, (e) Phase ii at $J_{\parallel}=-0.4$ and $H=0.2$, (f) Phase iii at $J_{\parallel}=-0.5$ and $H=1.5$, and (g) phase iv state at $J_{\parallel}=-1$ and $H=1.05$.
Black hexagons represent the first Brillouin zone.   
The middle and right two panels represent the data for layer B and layer C corresponding to the left two ones, respectively. 
}
\end{center}
\end{figure*}

\paragraph{Phase VIII: SkX with $(n_{\rm sk}^{\rm (A)}, n_{\rm sk}^{\rm (B)}, n_{\rm sk}^{\rm (C)})=(-1,x,-1)$.}
This phase appears in the small region next to Phase II, Phase III, Phase VII, and Phase V. 
As shown in Figs.~\ref{Fig:spin4}(b) and \ref{Fig:sq2}(a), the real-space spin configuration and the spin structure factor resemble those in Phase VII. 
Only the difference is found in the skyrmion number for layer B. 
In this Phase VIII, the skyrmion number takes non-integer values $0<x<2$. 
From the real-space picture, the sign of the $z$-spin component around the core denoted by the green circles in the middle panel of Fig.~\ref{Fig:spin4}(b) takes both positive and negative values. 
In other words, $S_i^z$ around the core shows the fluctuations with respect to the sign, which might be attributed to thermal fluctuations.  
This phase turns into Phase II (Phase VII) when $S_i^z>0$ ($S_i^z<0$) around all the cores. 
As the sign fluctuations around the core distinguish Phase II, Phase VII, and Phase VIII, a more careful analysis by using the finite-size scaling might be required, which is left in the future study.

\begin{figure*}[htb!]
\begin{center}
\includegraphics[width=1.0 \hsize]{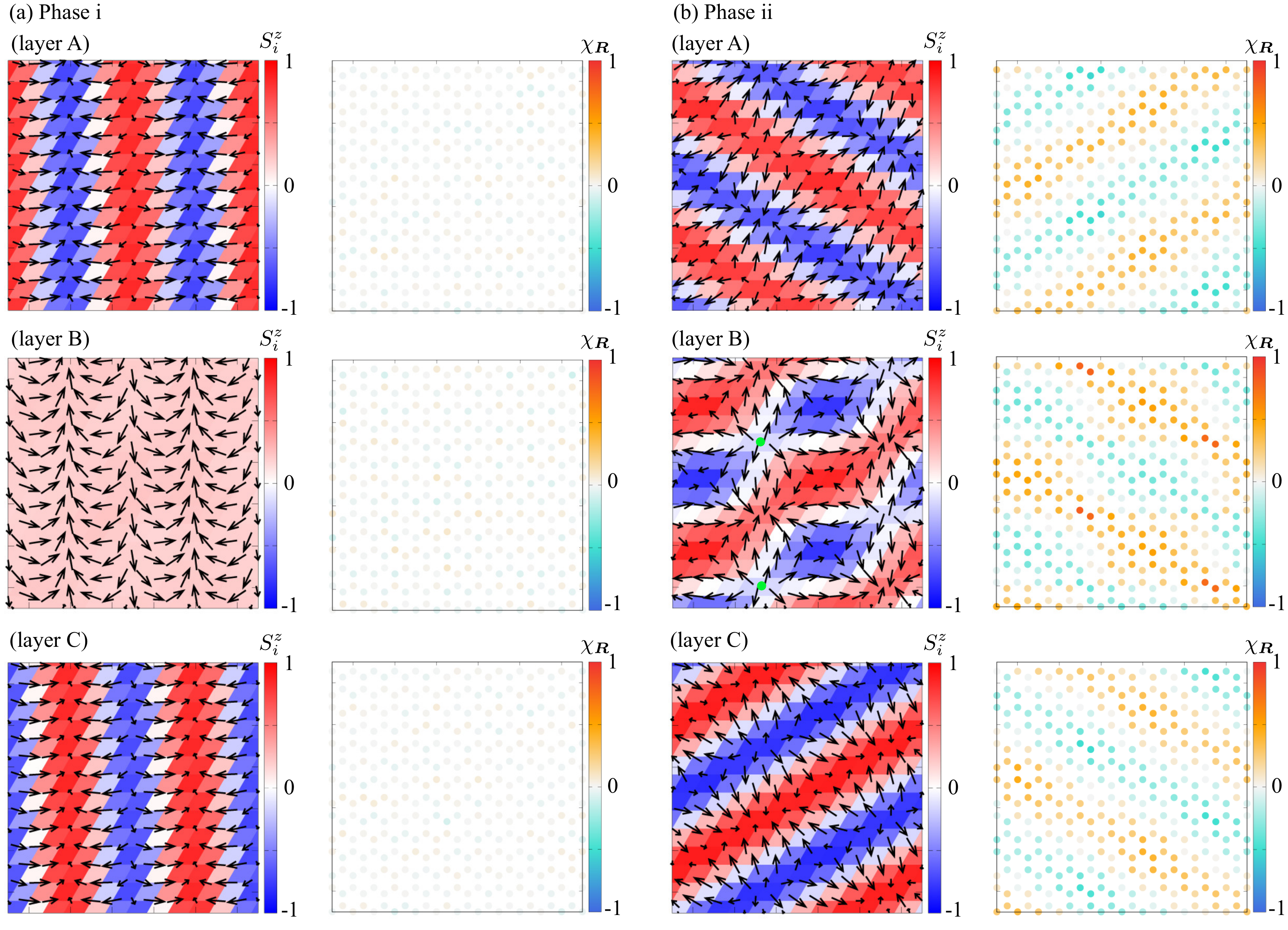} 
\caption{
\label{Fig:spin6}
Real-space spin (left panel) and scalar chiraity (right panel) configurations of (a) Phase i at $J_{\parallel}=-0.1$ and $H=0.4$ and (b) Phase ii at $J_{\parallel}=-0.4$ and $H=0.2$ on layer A (upper panel), layer B (middle panel), and layer C (lower panel).  
In the left panels, the arrows represent the $xy$ components of the spin moment and the color shows the $z$ component. 
The green circles represent the core positions discussed in the main text. 
}
\end{center}
\end{figure*}

\paragraph{Phase IX: SkX with $(n_{\rm sk}^{\rm (A)}, n_{\rm sk}^{\rm (B)}, n_{\rm sk}^{\rm (C)})=(0,1,0)$.}
This state is stabilized in both ferromagnetic and antiferromagnetic interlayer exchange interactions when $|J_{\parallel}|$ is relatively small in Fig.~\ref{fig: PD}(a). 
In contrast to the other SkX phases from Phase I to Phase VIII, the SkX spin texture appears only for layer B. 
For layer A, the spin configuration is characterized by the triple-$Q$ peak structure in the $xy$-spin component and the single-$Q$ peak structure in the $z$-spin component, as shown in Fig.~\ref{Fig:sq2}(b). 
Similar behavior of the spin structure factor is obtained for layer C, although the peak positions are located at different wave vectors; the dominant peak position at $\bm{Q}_2$ for layer A, while that at $\bm{Q}_3$ for layer C in Fig.~\ref{Fig:sq2}(b). 
Such a feature is also found in the real-space spin configuration for layers A and C in the left panel of Fig.~\ref{Fig:spin5}(a). 
These spin configurations accompany the chirality density wave; the dominant modulations are found along the $\bm{Q}_3$ ($\bm{Q}_2$) direction for layer A (layer C), as shown in the right panel of Fig.~\ref{Fig:spin5}(a). 
It is noted that a small uniform negative chirality occurs for both layers A and C. 

As shown in the middle panel of Fig.~\ref{Fig:spin5}(a), the spin texture for layer B is characterized by the anti-SkX with the skyrmion number of $+1$. 
This seems to be rather surprising, as the energy of the anti-SkX with a positive skyrmion number is usually higher than that of the SkX with a negative one in the triangular-lattice system with the threefold rotational symmetry, as discussed above. 
The emergence of the anti-SkX is presumably attributed to the effective threefold-symmetry-breaking field that arises from the anisotropic spiral spin textures for layers A and C. 
Indeed, the summation of the spin configuration over layers A and C leads to the same spin structure as layer B, which results in the energy gain by the ferromagnetic interlayer exchange interaction. 
Thus, the heterostructures sandwiched by the spiral states along the different directions are one of the ways of engineering the anti-SkX in the triangular-lattice system. 

A similar situation happens when the interlayer exchange interaction is antiferromagnetic $J_{\parallel} > 0$. 
We show the real-space spin configuration and spin structure factors at $J_{\parallel}=0.3$ and $H=0.4$ in Figs.~\ref{Fig:spin5}(b) and \ref{Fig:sq2}(c), respectively. 
In contrast to $J_{\parallel}<0$, the relative positions of the SkX core for layer B to layers A and C are different so as to have more anti-parallel spin components to gain the energy by the antiferromagnetic interlayer exchange interaction.

\begin{figure*}[htb!]
\begin{center}
\includegraphics[width=1.0 \hsize]{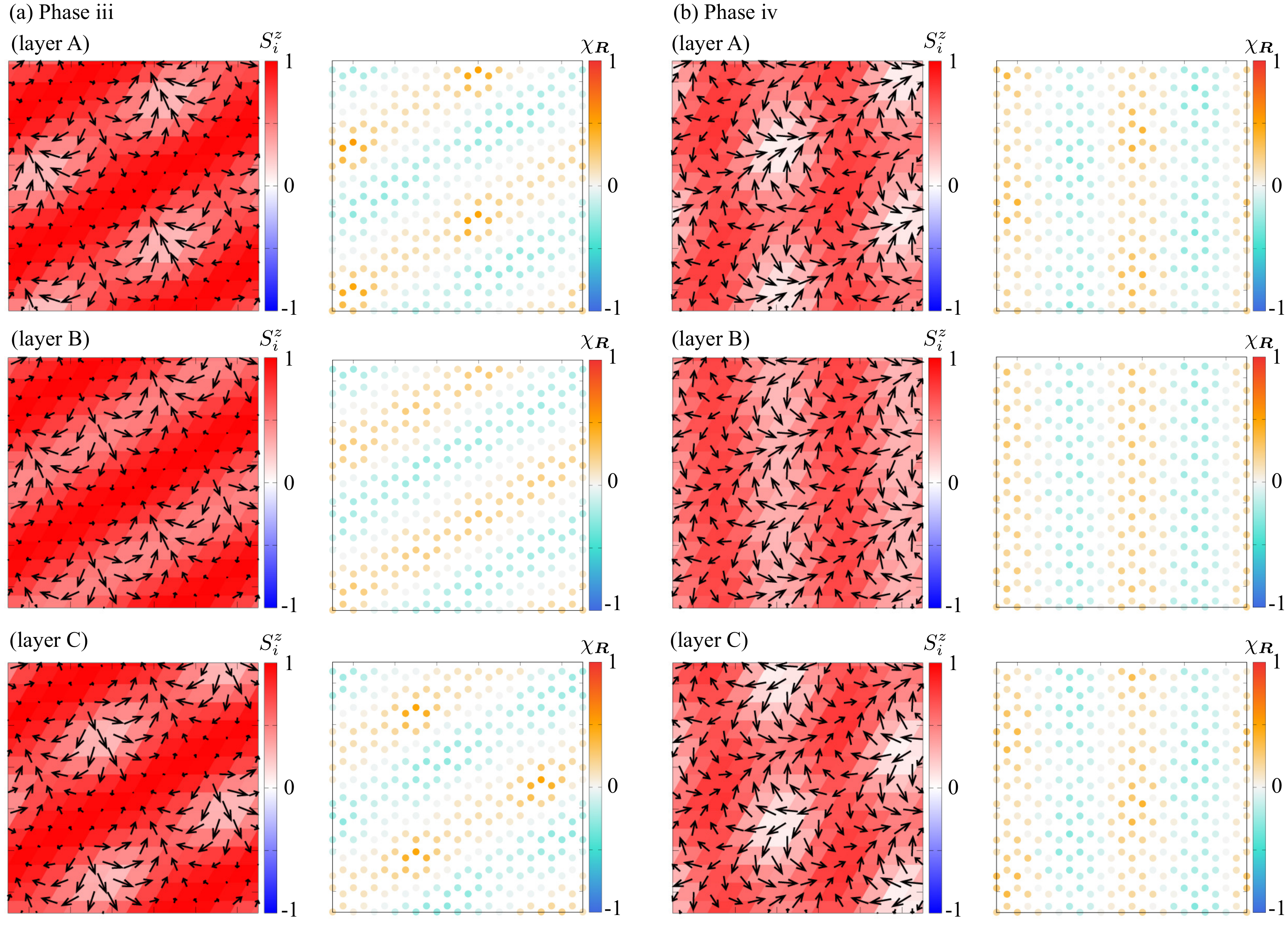} 
\caption{
\label{Fig:spin7}
Real-space spin (left panel) and scalar chiraity (right panel) configurations of (a) Phase iii at $J_{\parallel}=-0.5$ and $H=1.5$ and (b) Phase iv at $J_{\parallel}=-1$ and $H=1.05$ on layer A (upper panel), layer B (middle panel), and layer C (lower panel).  
In the left panels, the arrows represent the $xy$ components of the spin moment and the color shows the $z$ component.  
}
\end{center}
\end{figure*}

\paragraph{Phase i.}
This state appears in the three distinct regions in the phase diagram in Fig.~\ref{fig: PD}(a): One is the region for small $|J_{\parallel}|$ and small $H$, another is the region for large negative $J_{\parallel}$ and small $H$, and the other is the region for large positive $J_{\parallel}$.  
The spin configuration in these regions is mainly characterized by the single-$Q$ spiral wave for all the layers, where the type of the spiral waves for layers A and C is different from that for layer B, as shown in the case of $J_{\parallel}=-0.1$ in Fig.~\ref{Fig:spin6}(a). 
When considering $J_{\parallel}=0$, the spiral wave corresponds to the vertical spiral wave for layers A and C and the conical spiral for layer B, where the ordering vector $\bm{Q}_\nu$ is arbitrary in each layer. 
Here, the spiral plane for the vertical spiral state lies in the plane perpendicular to $\bm{D}_\nu^{(\eta)}$, while that for the conical state lies in the $xy$ plane. 
In the presence of $J_{\parallel}$, the dominant $\bm{Q}_\nu$ component becomes the same for three layers, as shown in Fig.~\ref{Fig:sq2}(d). 
In addition, the spiral plane for layers A and C is continuously tilted from the vertical spiral to the conical spiral, as shown in the upper and lower panels of Fig.~\ref{Fig:spin6}(a). 
Reflecting the single-$Q$ nature, the local chirality is suppressed for all the layers; a slightly staggered component for the upward and downward triangles appears under the conical spiral. 
As a result, this phase does not have a uniform scalar chirality. 

\paragraph{Phase ii.}
This phase is stabilized in both ferromagnetic and antiferromagnetic interlayer exchange interactions for small $|J_{\parallel}|$ and small $H$ shown in Fig.~\ref{fig: PD}(a). 
The spin configuration in this state in Fig.~\ref{Fig:spin6}(b) is similar to that in Phase IX in Fig.~\ref{Fig:spin5}(a) except for the following two points. 
One is the inequivalence between the spin configuration for layers A and C; in Phase ii, the $z$ component of the spin structure factor for layer A exhibits the double-$Q$ peak with different intensities, while that for layer C exhibits the single-$Q$ peak (the negligibly small intensity is found at the other $\bm{Q}_\nu$), as shown in Fig.~\ref{Fig:sq2}(e). 
On the other hand, both layers show the double-$Q$ peak with equal intensity in Phase IX as shown in Fig.~\ref{Fig:sq2}(b). 
The other is no skyrmion number for layer B. 
The zero skyrmion number for layer B is due to the quasi-stripe structure along the $\bm{Q}_2$ direction, as shown in the middle panel of Fig.~\ref{Fig:spin6}(b). 
Compared to the real-space spin configuration in Phase IX in Fig.~\ref{Fig:spin5}(a), one finds that the main difference appears in the sign of $S_i^z$ around the core denoted by the green circles in Fig.~\ref{Fig:spin6}(b), which results in the sign reversal of local scalar chirality. 
When its sign is reversed while increasing $H$, Phase ii turns into Phase IX. 
Although there is no skyrmion number in Phase ii, the uniform scalar chirality arises, as shown in Figs.~\ref{Fig: Mag1} and \ref{Fig: Mag2}.

\paragraph{Phase iii.}
This phase appears as a stable state in the high-field region in the phase diagram in Fig.~\ref{fig: PD}(a). 
For layers A and C, the spin configuration is characterized by the anisotropic triple-$Q$ peak structure in both $xy$- and $z$-spin components; the dominant peak positions are located at $\bm{Q}_1$ and $\bm{Q}_3$ in the $xy$-spin component, while those are at $\bm{Q}_2$ in the $z$-spin component, as shown in Fig.~\ref{Fig:sq2}(f). 
Meanwhile, the double-$Q$ (single-$Q$) peak appears in the $xy$($z$)-spin component for layer B. 
This spin state accompanies the chirality density waves along the $\bm{Q}_2$ direction as shown in the right panel of Fig.~\ref{Fig:spin7}(a). 
There is a negative small scalar chirality in this phase. 
It is noted that the other high-field phase with the triple-$Q$ peak in the $xy$-spin component and no peak in the $z$-spin component at $\bm{Q}_\nu$ appears for small $|J_{\parallel}|$  in the vicinity of the fully-polarized state as an almost energetically-degenerate state. As it is difficult to distinguish between them, we summarize them as Phase iii. 

\paragraph{Phase iv.}
This phase appears next to Phase iii upon decreasing $H$ for large $|J_{\parallel}|$ as shown in Fig.~\ref{fig: PD}(a). 
The real-space spin and scalar chirality configurations in Phase iv are similar to those in Phase iii, as shown in Figs.~\ref{Fig:spin7}(a) and \ref{Fig:spin7}(b). 
Their difference is clearly found in the spin structure factor, as shown in Figs.~\ref{Fig:sq2}(f) and \ref{Fig:sq2}(g); the intensities in the $xy$-spin component at $\bm{Q}_\nu$ are different for all the layers in Phase iv. 
This state also exhibits nonzero uniform scalar chirality.

\subsection{Magnetic-field dependence}
\label{sec: Magnetic-field dependence}

\begin{figure}[htb!]
\begin{center}
\includegraphics[width=1.0 \hsize]{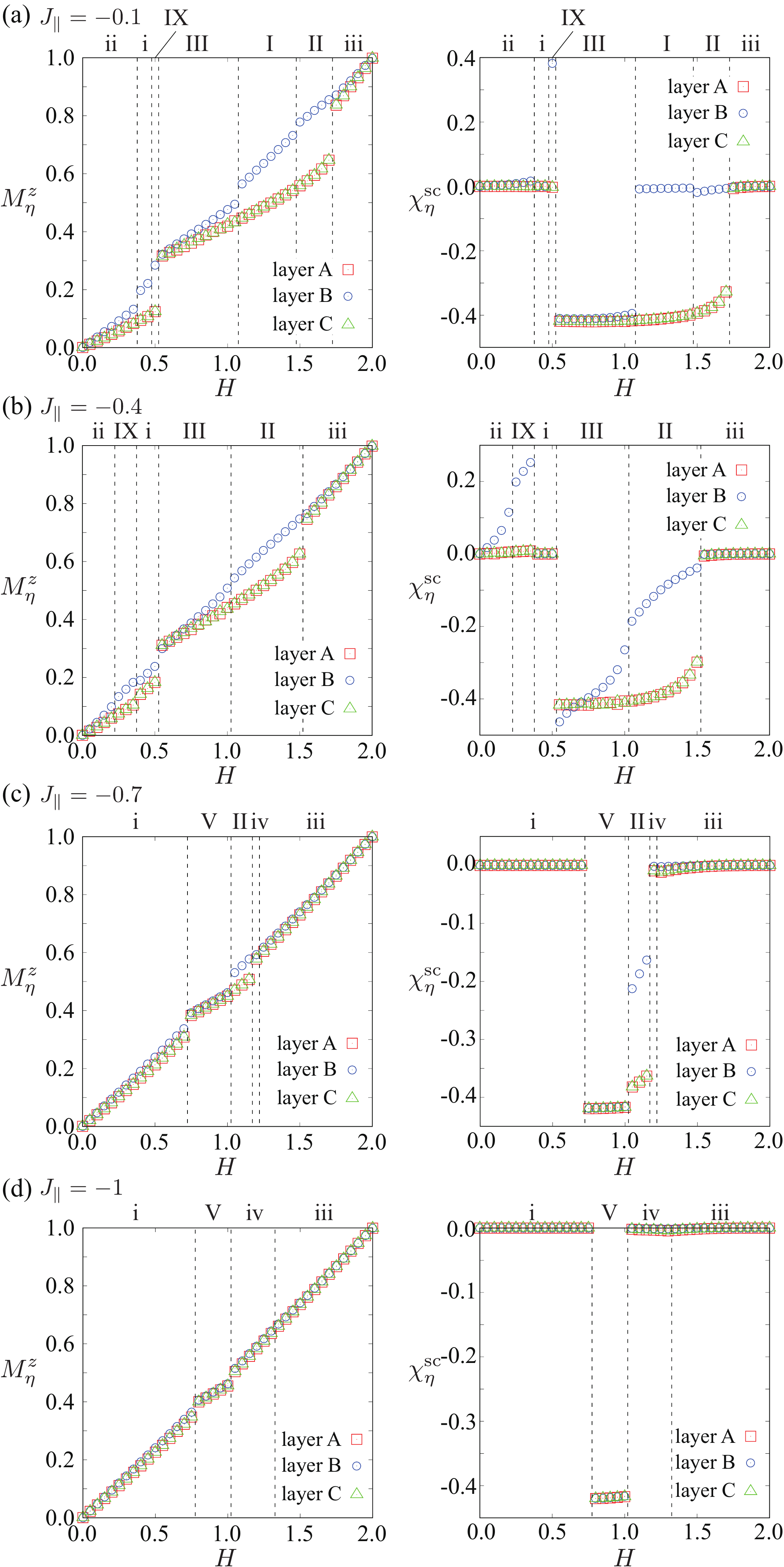} 
\caption{
\label{Fig: Mag1}
$H$ dependences of the magnetization $M^{z}_\eta$ (left panel) and the scalar chirality $\chi^{\rm sc}_{\eta}$ (right panel) for layers $\eta=$ A, B, and C at (a) $J_{\parallel}=-0.1$, (b) $J_{\parallel}=-0.4$, (c) $J_{\parallel}=-0.7$, and (d) $J_{\parallel}=-1$. 
The vertical dashed lines represent the phase boundaries between different phases, where the corresponding phases are presented above in each figure. 
}
\end{center}
\end{figure}

\begin{figure}[htb!]
\begin{center}
\includegraphics[width=1.0 \hsize]{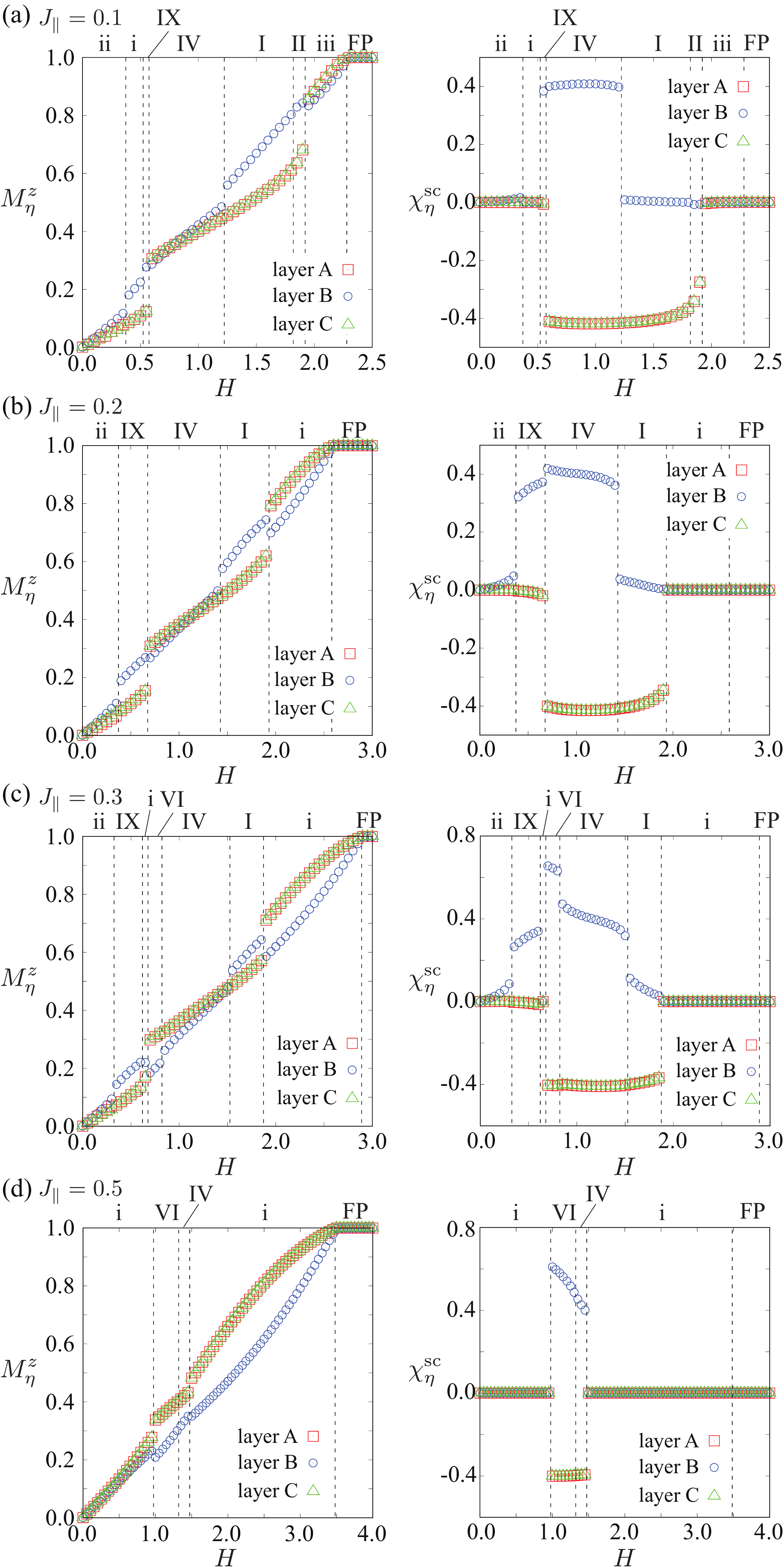} 
\caption{
\label{Fig: Mag2}
$H$ dependences of the magnetization $M^{z}_\eta$ (left panel) and the scalar chirality $\chi^{\rm sc}_{\eta}$ (right panel) for layers $\eta=$ A, B, and C at (a) $J_{\parallel}=0.1$, (b) $J_{\parallel}=0.2$, (c) $J_{\parallel}=0.3$, and (d) $J_{\parallel}=0.5$. 
The vertical dashed lines represent the phase boundaries between different phases, where the corresponding phases are presented above in each figure. 
}
\end{center}
\end{figure}
 
We discuss the phase sequence for several values of $J_{\parallel}$ against $H$. 
Figure~\ref{Fig: Mag1} shows the $H$ dependences of the magnetization $M_\eta^z$ and the scalar chirality $\chi_\eta^{\rm sc}$ in each layer under the ferromagnetic interlayer exchange interaction. 
The data correspond to $J_{\parallel}=-0.1$ in Fig.~\ref{Fig: Mag1}(a), $J_{\parallel}=-0.4$ in Fig.~\ref{Fig: Mag1}(b), $J_{\parallel}=-0.7$ in Fig.~\ref{Fig: Mag1}(c), and $J_{\parallel}=-1$ in Fig.~\ref{Fig: Mag1}(d). 
As shown in Figs.~\ref{Fig: Mag1}(a)-\ref{Fig: Mag1}(d), the almost phase transitions are characterized by the first-order transitions with the jumps of $M_\eta^z$ and $\chi_\eta^{\rm sc}$. 
Among them, the phase transitions between Phase ii and Phase IX in Fig.~\ref{Fig: Mag1}(b), Phase III and Phase II in Fig.~\ref{Fig: Mag1}(b), and Phase iv and Phase iii in Figs.~\ref{Fig: Mag1}(c) and \ref{Fig: Mag1}(d) as well as the phase transition between the fully-polarized state and Phase iii seem to be the second-order phase transitions, where spin- and chirality-related quantities continuously change at the transition. 
In the case of the transition between Phase ii and Phase IX, the real-space spin configuration in Phase ii is transformed into that in Phase IX by reversing the sign of $S_i^z$ around the core denoted by the green circles for layer B in Fig.~\ref{Fig:spin6}(b), as discussed in Sec.~\ref{sec: Details of magnetic phases}. 
For the transition between Phase II and Phase III, the spin configuration in Phase III turns into that in Phase II by reversing the sign of $S_i^z$ around the skyrmion core denoted by the green circles for layer B in Fig.~\ref{Fig:spin2}(a). 
For the transition between Phase iii and Phase iv, they are transformed with each other when changing the $xy$ component of the spin in the double-$Q$ peak for all the layers; the dominant double-$Q$ peaks with the same intensity correspond to the Phase iii, while those with the different intensity corresponds to Phase iv, as shown in Figs.~\ref{Fig:sq2}(f) and \ref{Fig:sq2}(g). 

Figure~\ref{Fig: Mag2} shows the results for the antiferromagnetic interlayer exchange interaction: $J_{\parallel}=0.1$ in Fig.~\ref{Fig: Mag2}(a), $J_{\parallel}=0.2$ in Fig.~\ref{Fig: Mag2}(b), $J_{\parallel}=0.3$ in Fig.~\ref{Fig: Mag2}(c), and $J_{\parallel}=0.5$ in Fig.~\ref{Fig: Mag2}(d). 
In this case, the transition between Phase VI and Phase IV in Fig.~\ref{Fig: Mag2}(d) is of second order as well as the transitions between the fully-polarized state and Phase i (or Phase iii). 
For this transition, the type of the constituent waves in the multiple-$Q$ spin configuration for layer B changes; the triple-$Q$ sinusoidal waves in Phase VI turn into the triple-$Q$ spiral waves in Phase IV by changing the relative angle between the $xy$ and $z$ spins~\cite{yambe2021skyrmion}.

\section{Summary}
\label{sec: Summary}

To summarize, we have investigated the instability toward the SkX in the centrosymmetric multi-layer system. 
By focusing on the layer-dependent DM interaction in the trilayer triangular-lattice structure, we found multifarious SkX phases depending on the interlayer exchange interaction and the magnetic field. 
The phase diagram was constructed by performing the simulated annealing for the effective spin model. 
As a result, we obtained nine types of SkX phases, which are characterized by different multiple-$Q$ superpositions, scalar chirality distributions, and skyrmion numbers. 
In particular, we showed that the middle layer without the DM interaction exhibits multiple skyrmion numbers from $-2$ to $+2$ depending on the model parameters. 
This indicates that the layered system with the layer-dependent DM interaction is promising to realize a variety of the SkXs, such as the twisted SkXs, the anti-SkXs, and the high-topological-number SkXs, which have not been stabilized by the polar-type DM interaction in the single-layer system. 

Although we focus on the layer degree of freedom, a similar situation is expected in the systems where the inversion symmetry is preserved globally but broken intrinsically at local sites so that the sublattice-dependent DM interaction is present. 
Such a situation have been found in various lattice systems with the sublattice degree of freedom, such as the zigzag~\cite{Yanase_JPSJ.83.014703,Hayami_doi:10.7566/JPSJ.84.064717,Hayami_doi:10.7566/JPSJ.85.053705,Sumita_PhysRevB.93.224507,cysne2021orbital,Suzuki_PhysRevB.105.075201,yatsushiro2021microscopic}, honeycomb~\cite{Kane_PhysRevLett.95.226801,Hayami_PhysRevB.90.081115,yanagi2017optical,Yanagi_PhysRevB.97.020404}, and diamond~\cite{Fu_PhysRevLett.98.106803,Hayami_PhysRevB.97.024414,Ishitobi_doi:10.7566/JPSJ.88.063708} structures. 
Indeed, the fractional antiferromagnetic SkX has recently been observed in MnSc$_2$S$_4$ with the diamond structure~\cite{Gao2016Spiral,gao2020fractional}. 
Our results provide a possibility of realizing further exotic SkXs in the systems with the layer/sublattice-dependent DM interaction.

\appendix
\section{Case of different DM interactions}
\label{sec: app}

\begin{figure*}[htb!]
\begin{center}
\includegraphics[width=1.0 \hsize]{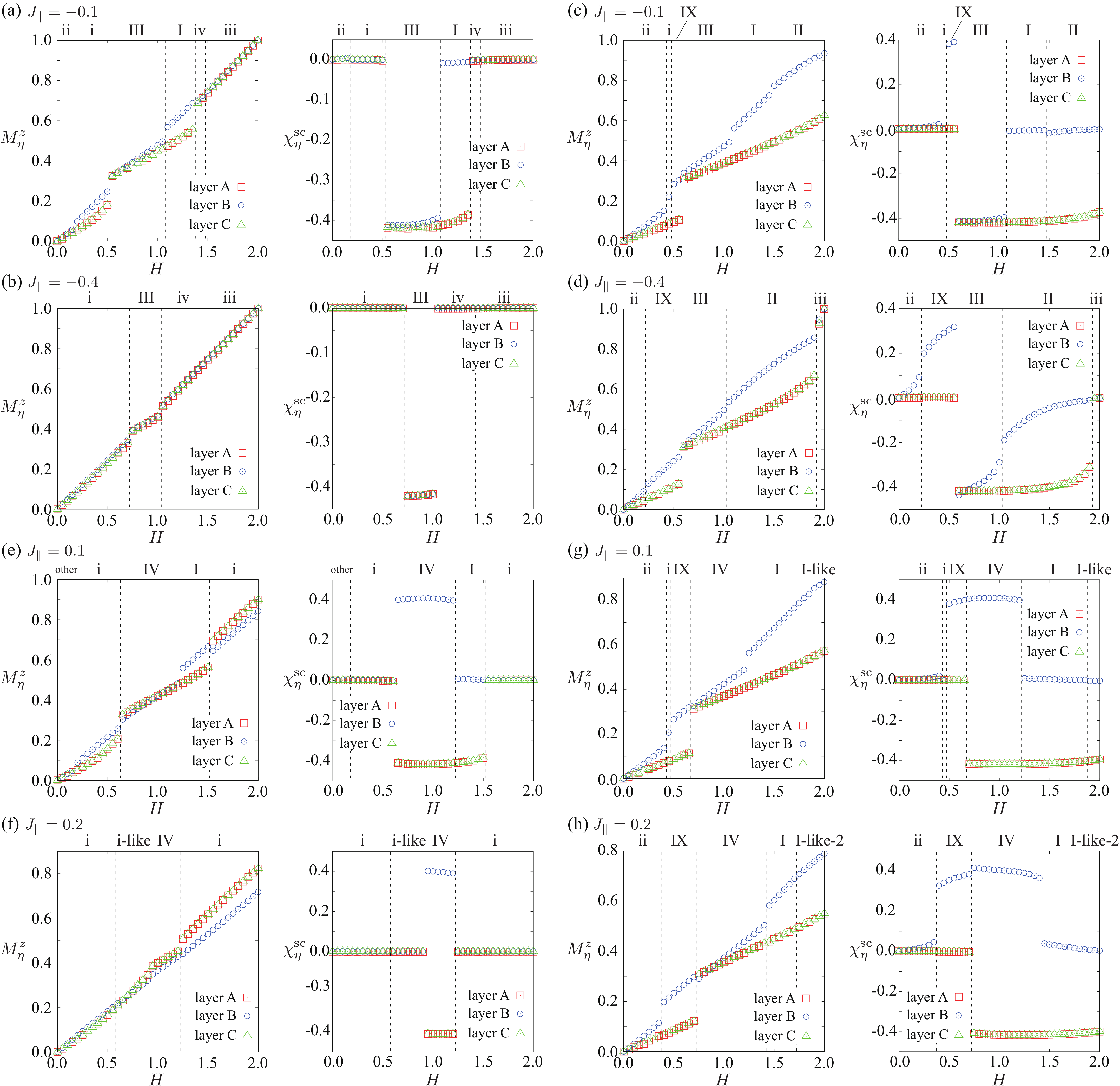} 
\caption{
\label{Fig: Mag_app}
$H$ dependences of the magnetization $M^{z}_\eta$ (left panel) and the scalar chirality $\chi^{\rm sc}_{\eta}$ (right panel) for layers $\eta=$ A, B, and C at (a) $J_{\parallel}=-0.1$ and $D=0.1$, (b) $J_{\parallel}=-0.4$ and $D=0.4$, (c) $J_{\parallel}=0.1$ and $D=0.1$, and (d) $J_{\parallel}=0.2$ and $D=0.4$. 
The vertical dashed lines represent the phase boundaries between different phases, where the corresponding phases are presented above in each figure. 
}
\end{center}
\end{figure*}

In this Appendix, we briefly discuss the results for different $D$. 
Figure~\ref{Fig: Mag_app} shows the $H$ dependences of $M^{z}_\eta$ and $\chi^{\rm sc}_{\eta}$ for layers $\eta=$ A, B, and C at $D=0.1$ [Figs.~\ref{Fig: Mag_app}(a), \ref{Fig: Mag_app}(b), \ref{Fig: Mag_app}(e), and \ref{Fig: Mag_app}(f)] and $D=0.4$ [Figs.~\ref{Fig: Mag_app}(c), \ref{Fig: Mag_app}(d), \ref{Fig: Mag_app}(g), and \ref{Fig: Mag_app}(h)] for different $J_{\parallel}$. 
The phases are presented above in each figure. 
In Figs.~\ref{Fig: Mag_app}(f)-\ref{Fig: Mag_app}(h), ``I-like", ``I-like-2", and ``i-like IV" correspond to the same layer-dependent skyrmion number as Phase I, Phase I, and Phase i, respectively, but their spin configurations are slightly different. 
For example, in the spin configuration denoted by ``I-like", the $xy$-spin component has the same intensities as $S^{xy}_{\rm B}(\bm{Q}_1)=S^{xy}_{\rm B}(\bm{Q}_2)=S^{xy}_{\rm B}(\bm{Q}_3)$ but the $z$-spin component do no have a peak structure at $\bm{Q}_\nu$, which differs from Phase I. 
In Fig.~\ref{Fig: Mag_app}(e), ``other" represents a different triple-$Q$ state without a nonzero skyrmion number.

For both cases of $J_{\parallel}<0$ and $J_{\parallel}>0$, the SkX phases become more (less) stabilized for large (small) $D$. 
Meanwhile, we find that Phase IX and Phase II do not appear for small $D$, as shown in Figs.~\ref{Fig: Mag_app}(a), \ref{Fig: Mag_app}(b), \ref{Fig: Mag_app}(e), and \ref{Fig: Mag_app}(f), where the skyrmion number for layers A and C is different from that for layer B.  
This result indicates that the overall tendency in terms of the stabilization is similar among the different types of the SkXs while changing $D$, but the SkX phases consisting of layers with different skyrmion numbers tend to be destabilized for small $D$. 
In other words, the layer-dependent DM interaction is essentially important for the stabilization of such complicated SkX phases.

\begin{acknowledgments}
This research was supported by JSPS KAKENHI Grants Numbers JP19K03752, JP19H01834, JP21H01037, and by JST PRESTO (JPMJPR20L8).
Parts of the numerical calculations were performed in the supercomputing systems in ISSP, the University of Tokyo.
\end{acknowledgments}

\bibliographystyle{apsrev}
\bibliography{ref}

\end{document}